\documentstyle[preprint, eqsecnum, aps]{revtex}

\def\displayfrac#1#2{\frac{\displaystyle #1}{\displaystyle #2}}

\begin{document}
\title{Wigner distributions and quantum mechanics on Lie groups: the case of
  the regular representation}
\author{N. Mukunda\thanks{email: nmukunda@cts.iisc.ernet.in}}
\address{Centre for Theoretical
Studies, Indian Institute of Science, {Bangalore~560~012,} India\\
and\\
Jawaharlal Nehru Centre for Advanced Scientific Research,
 Jakkur, Bangalore~560~064, India}
\author{ Arvind\thanks{Present Address: Department of Physics,
Carnegie Mellon University, Pittsburgh  PA 15213 USA}
\thanks{email:~ xarvind@andrew.cmu.edu}} 
\address{Department of Physics, Guru Nanak Dev University, Amritsar 143005, 
India}
\author{S. Chaturvedi\thanks{e-mail: scsp@uohyd.ernet.in}} 
\address{ School of Physics, University of Hyderabad, Hyderabad 500 046,
India}
\author{ R.Simon\thanks{email: simon@imsc.res.in}}
\address{The Institute of 
 Mathematical Sciences, C. I. T. Campus, Chennai 600 113, India}
 \date{\today}
 \maketitle

\begin{abstract}
We consider the problem of setting up the Wigner distribution for states of a
quantum system whose configuration space is a Lie group. The basic properties
of Wigner distributions in the familiar Cartesian case are systematically
generalised to accommodate new features which arise when the configuration
space changes from $n$-dimensional Euclidean space ${\cal R}^n$ to a Lie group 
$G$. The notion of canonical momentum is carefully analysed, and the meanings
of marginal probability distributions and their recovery from the Wigner
distribution are clarified. For the case of compact $G$ an explicit definition
of the Wigner distribution is proposed, possessing all the required
properties. Geodesic curves in $G$ which help introduce a notion of 
the `mid point' of two group elements play a central role in the construction. 
\end{abstract}
\section{Introduction}

The method of Wigner distributions \cite{1} as a description of states of
quantum mechanical systems appeared in 1932, quite early in the
history of quantum mechanics.  For systems whose kinematics is
based upon a set of Heisenberg canonical commutation relations,
it gives a way of describing both pure and mixed states in a
classical phase space setting, at the level of density
operators.  Thus it must be sharply distinguished in
mathematical structure from the Hilbert space state vector or
wave function description of states, which highlights the
superposition principle of quantum mechanics.  In the Wigner
distribution language, this principle is not obvious or
manifest, but is somewhat hidden in the formalism.  On the other
hand, the formation of convex classical  statistical mixtures of
general states to generate new states becomes much more obvious.
Somewhat later it was appreciated that the Wigner distribution
way of describing quantum mechanical states is dual to, or is
naturally accompanied by, the Weyl ordering rule \cite{2} - a convention
by which one can set up a one-to-one correspondence between
operators in quantum theory (in the case of the Heisenberg
commutation relations) and c-number dynamical phase space
variables for the comparison classical system.  Thus expectation
values for general quantum dynamical  variables in general
quantum states can be faithfully expressed in the full
operator-state vector language, or equally well in a completely
c-number classical phase space language.  In this general scheme
, the group $Sp(2n, R)$ of linear canonical transformations for
$n$ Cartesian degrees of freedom, and the related two-fold
covering group $Mp(2n)$, play prominent roles \cite{2a}.

An important property of the Wigner distribution for a general
quantum state is that while it is a real function on the
classical phase space, it is not always pointwise nonnegative.
Therefore it is usually called a quasi probability distribution,
and can not be interpreted as a phase space probability density
in the sense of classical statistical mechanics.  However, the
Wigner distribution does have the attractive property that the
marginal distributions, obtained by integrating away either the
momentum or the position variables, do reproduce the correct
nonnegative position space and momentum space probability
distributions respectively as specified by quantum mechanics.
This recovery or reproducibility of correct marginal
distributions is of course maintained  even after unitary
action by any $Mp(2n)$ element.

There have been several attempts \cite{3}-\cite{12b} over the years to 
generalise the method of Wigner distributions to handle quantum mechanical
situations where the basic kinematics is defined, not by
Heisenberg type canonical commutation relations, but by some Lie
group which acts as the covariance group of the system of
interest.  (As will become clear in the sequel, the traditional
case is also governed by a group, namely the Abelian group
${\cal R}^n$ of translations in $n$-dimensional Euclidean
space).  A commonly studied group is $SU(2)$, in the context of
spin systems as well as two-level atoms. One of the important 
early efforts at providing a general group theoretical setting 
for the Wigner distribution is due to Stratonovich \cite{alpha}. 
In this context we should also mention the comprehensive monograph of 
Dubin et al \cite{beta}.  It seems to us, however, that in most
of these attempts the requirement that certain marginal probability
distributions be recovered in a natural way from the Wigner distribution
corresponding to a general quantum state, which as mentioned above is an
important feature of the usual Cartesian case, is not discussed in a
satisfactory manner; in some of these works this important aspect is not 
considered at all. 

The aim of the present paper is to develop from first principles  the basic
features of quantum kinematics for a system whose configuration space is a
general non Abelian Lie group $G$ rather than a Cartesian space ${\cal R}^n$,
and then set up a corresponding Wigner distribution formalism which respects
the requirement that natural marginal probability distributions are reproduced
in a simple manner. This involves several extensions or modifications of the
familiar formalism in the Cartesian case.  The role of Schr\"odinger wave 
functions is of course 
now played by complex square integrable functions on $G$, and after 
normalisation each such wave function determines a probability distribution 
over $G$. The meaning and definition of canonical momentum variables, and 
determination of the momentum space probability distribution for a given 
state, are however nontrivial questions in which the many structural features
associated with $G$ play important roles. In particular for a non Abelian $G$
canonical momenta in quantum theory become non commuting operators, leading to
deep changes in the meaning of momentum eigenstates, momentum eigenvalues and
momentum space etc.. It is here that the unitary representation theory of $G$
plays an important role. We show that all these features can be properly taken
into account, and a fully satisfactory Wigner distribution can be set up as a
function of carefully chosen arguments. This turns out to have simple
transformation properties under $G$ action and also to reproduce the marginal
`position' and `momentum' space probability distributions properly. 

It needs to be emphasised that our interest in  developing a Wigner
distribution formalism for systems whose configuration space is a non Abelian
group is not purely  academic. In fact, there  many familiar systems which 
fall in this category. A general rigid body which has the group $SO(3)$ as 
its configuration space is a case in point. Another well studied example 
in this category is the relativistic spherical top \cite{HR} whose 
configuration space is the non compact group $SO(3,1)$. 

The material of this paper is arranged as follows. In Section II we recall the
main definitions and properties of Wigner distributions in the Cartesian
situation. We emphasise several familiar features in this case: the
possibility of use of the classical phase space as the domain of definition of
Wigner distributions; the roles of the groups $Sp(2n,R)$ and $Mp(2n)$; the 
reality but in general loss of pointwise positivity of Wigner distributions; 
and the recovery of the coordinate space and momentum space probability 
distributions for a given state by integrating over  half the arguments of
the Wigner distribution. Section III describes briefly the properties of 
Wigner distributions in the angle-angular momentum case \cite{13}. 
This brings out some new features, namely loss of the classical phase space 
as the domain of definition of Wigner distributions, and absence 
of replacements for the groups $Sp(2n,R)$ and $Mp(2n)$, which indicate the 
type of changes we should expect in the case of a general Lie group. 
Section IV analyses in some detail the classical phase space that goes with a 
non Abelian Lie group $G$ as configuration space. Both global intrinsic and 
local coordinate based descriptions are given, and the associated classical 
Poisson Bracket relations are developed and described in several ways. 
In particular a careful analysis of the concept of classical canonical  
momenta in this case is provided. The transition to quantum mechanics, based 
on Schr\"odinger wave functions over $G$, is then outlined. It is emphasised  
that a naive generalisation of the usual canonical Heisenberg commutation 
relations is not possible, and all the concepts of position operators, 
momentum operators and their commutation relations have to be treated with 
care. A brief Section V indicates the kinds of new features we may expect 
to appear, based on the results and discussions of Sections III and IV.  
In Section VI we pose the main problem of defining Wigner distributions in a 
suitable way, with suitable choice of arguments, subject to the main 
requirements already mentioned above: reasonable transformation laws under 
$G$ action, recovery of marginal probability distributions, and capturing the 
full information contained in a general pure or mixed quantum state. We 
propose a solution to this problem, possessing all the desired properties. 
We find that our solution uses in an essential and interesting manner the 
concept and properties of geodesics in $G$ leading to the notion of 
a `mid point' of two group elements, a key ingredient in our construction. 
For definiteness we confine ourselves to the case of compact $G$. 
Section VII shows how the known results of Sections II and III, for Cartesian 
quantum mechanics and for the angle angular- momentum pair, are easily 
recovered from the general case. They correspond actually to the choices 
$G= {\cal R}^n$ and $G=SO(2)$, which are both Abelian and respectively 
non compact and compact. The case of $G=SU(2)$ is then briefly considered, 
giving adequate background details so that the structure of the 
Wigner distribution can be easily appreciated. Some of the important 
differences compared to the Cartesian case, as well as to earlier 
approaches, are mentioned. Section VIII contains some concluding remarks. We
have included two appendices. Appendix A recollects basic results from the
theory of the regular representation of $G$ in the compact case, based
essentially on the Peter-Weyl theorem. In addition certain useful operator 
structures are set up, which help us understand better the construction of
Wigner distributions in Section VI. Appendix B discusses the question of
completeness of the information content in the Wigner distribution set up in  
Section VI, and generalisations of the Weyl exponential operators to the non 
Abelian Lie group case.      
\section{The Wigner distribution in the Cartesian case}

It is useful to recall briefly the usual definition and the
basic properties of the Wigner distribution in the case of
Cartesian quantum mechanics, and to highlight those important
features which are likely to need generalisation when we later
take up the treatment of quantum mechanics on a general Lie
group.

We consider a quantum system whose kinematics is based on $2n$
hermitian irreducible Cartesian position and momentum operators
$\hat{q}_r, \hat{p}_r, r=1,2,\ldots, n$, obeying the standard
Heisenberg commutation relations 
\begin{eqnarray}
[\hat{q}_r,\hat{p}_s] = i \hbar \delta_{rs},\;\; [\hat{q}_r,
\hat{q}_s] = [\hat{p}_r, \hat{p}_s] = 0,\;\; r,s=1,2,\ldots,n.
\label{1}
\end{eqnarray}

\noindent
It is useful to express these relations more compactly by
defining a $2n$-dimensional column vector with hermitian
operator entries, 
\begin{eqnarray}
\hat{\xi}=\left(\hat{\xi}_a\right) =
\left(\hat{q}_1\ldots\hat{q}_n\;\;
\hat{p}_1\ldots\hat{p}_n\right)^T,\; a=1,2,\ldots, 2n ,
\label{2}
\end{eqnarray}

\noindent
and a real antisymmetric nondegenerate $2n$ dimensional
symplectic metric matrix $\beta$ as 
\begin{eqnarray} \beta =
\left(\matrix{ 0_{n\times n}&1_{n\times n}\cr
-1_{n\times n} &0_{n\times n}}\right).  
\label{3}
\end{eqnarray}

\noindent 
Then eqn. $(\ref{1})$ can be written as 
\begin{eqnarray}
\left[\hat{\xi}_a, \hat{\xi}_b \right] = i \hbar \beta_{ab}
\label{4}
\end{eqnarray}

\noindent
These commutation relations and the hermiticity properties are
preserved when we subject the operators $\hat{\xi}_a$ to a real
linear transformation by any matrix of the symplectic group
$Sp(2n,R)$: 
\begin{mathletters} 
\begin{eqnarray} Sp(2n,R) &=&
\left\{S = 2n\times 2n \;\mbox{real matrix}\;| S           \beta
S^T = \beta\right\};\\ S~\in ~Sp(2n,R) &:& \hat{\xi}_a
\rightarrow \hat{\xi}^{\prime}_a = S_{ab} \hat{\xi}_b
,\nonumber\\ &&[\hat{\xi}^{\prime}_a , \hat{\xi}^{\prime}_b] =
i\;\hbar\;\beta_{ab}.  
\end{eqnarray} 
\end{mathletters}

\noindent
On account of the Stone-von Neumann theorem, such linear
transformations must be unitarily induced; i.e., for each
$S~\in~ Sp(2n,R)$, there exists a unitary operator
$\overline{U}(S)$, determined upto a phase, such that
\begin{eqnarray} \hat{\xi}^{\prime}_a = S_{ab} \hat{\xi}_b =
\overline{U}(S)^{-1} \hat{\xi}_a \overline{U}(S).
\end{eqnarray}

\noindent
These unitary operators give a unitary representation of
$Sp(2n,R)$ upto phases which cannot be totally eliminated, but
can at best be reduced to a sign ambiguity: \begin{eqnarray}
S^{\prime}, S~\in ~Sp(2n,R) : \overline{U}(S^{\prime})
\overline{U}(S) = \pm \overline{U}(S^{\prime}S).  \end{eqnarray}

\noindent
This situation may be expressed by the statement that one is
actually dealing here with a true representation of the group
$Mp(2n)$ which is a double cover of $Sp(2n,R)$.  These objects
will be seen to play important roles in the theory of Wigner
distributions in the present case.

Vectors in the Hilbert space ${\cal H}$ on which the
$\hat{\xi}_a$ are irreducibly represented may be described by
their Schr\"{o}dinger wave functions in the usual manner:
\begin{eqnarray} |\psi\rangle\in {\cal H} :
\psi(\underline{q}) &=& \langle\underline{q}|\psi\rangle
,\nonumber\\ \langle\underline{q}^{\prime}|\underline{q}\rangle
&=& \delta^{(n)}\left(\underline{q}^{\prime}-
\underline{q}\right),\nonumber\\ \langle\psi|\psi\rangle =
\parallel\psi\parallel^2 &=& \int\limits_{{\cal R}^n} d^n q
|\psi(\underline{q})|^2.  \end{eqnarray}

\noindent
The (ideal) kets $|\underline{q}\rangle$ are simultaneous
eigenvectors of the $n$ commuting position operators $\hat{q}_r,
r=1,\ldots, n$.  Alternatively we may describe them by their
momentum space wave functions $\tilde{\psi}(\underline{p})$ by
taking the overlap with simultaneous eigenvectors of the
commuting momentum operators  $\hat{p}_r,r=1,\ldots, n$:
          
     \begin{eqnarray} \tilde{\psi}(\underline{p}) &=&
\langle\underline{p}|\psi\rangle = \int\limits_{{\cal R}^n}
\displayfrac{d^n q}{(2\pi\hbar)^{n/2}}
\psi(\underline{q})\exp(-i\;\underline{p}\cdot\underline{q}/\hbar),
\nonumber\\ \langle\underline{q}|\underline{p}\rangle
&=&(2\pi\hbar)^{-n/2} \exp(i\;\underline{q}\cdot\underline{p}
/\hbar),\nonumber\\ \langle\psi|\psi\rangle &=&
\int\limits_{{\cal R}^n} d^np
|\tilde{\psi}(\underline{p})|^2.  \end{eqnarray}

Given a pure state $|\psi\rangle$ of the above quantum system,
the corresponding Wigner distribution is a function
$W(\underline{q}, \underline{p})$ of $2n$ classical real
variables, ie., a function on ${\cal R}^{2n}$.  In analogy with
eqn.$(\ref{2})$ we assemble the arguments  $q_1\ldots q_n\; p_1\ldots
p_n$ into a real $2n$-component column vector
$\xi=(\xi_a)=(q_1\ldots q_n\;p_1\ldots p_n)^T$, and then
$W(\xi)$ is defined by a partial Fourier transformation:
\begin{eqnarray} W(\xi)=(2\pi\hbar)^{-n} \int\limits_{{\cal
R}^n} d^nq^{\prime}
\psi\left(\underline{q}-\frac{1}{2}\underline{q}^{\prime}\right)
\psi\left(\underline{q}+\frac{1}{2}\underline{q}^{\prime}\right)^*
\exp\left(i\;\underline{p}\cdot\underline{q}^{\prime}/\hbar\right).
\label{10}
\end{eqnarray}

\noindent
Here the dependence of $W(\xi)$ on $\psi$ is left implicit.  For
a general mixed state we define $W(\xi)$ through the
configuration space matrix elements of the density operator
$\hat{\rho}$: 
\begin{eqnarray}
W(\xi)=(2\pi\hbar)^{-n}\int\limits_{{\cal R}^n} d^n q^{\prime}
\left\langle\underline{q}-\frac{1}{2}
\underline{q}^{\prime}\big| \hat{\rho}\big|\underline{q} +
\frac{1}{2}\underline{q}^{\prime} \right\rangle
\exp\left(i\;\underline{p}\cdot\underline{q}^{\prime}
\big/\hbar\right) , 
\label{11}
\end{eqnarray}

\noindent
once again leaving the dependence on $\hat{\rho}$ implicit.  It
is clear by construction that $W(\xi)$ is a real phase space
function.  The recovery of the proper nonnegative marginal
probability distributions is demonstrated by 
\begin{eqnarray}
\int\limits_{{\cal R}^n} d^n p\; W(\xi) &=&
|\psi(\underline{q})|^2, \nonumber\\ \int\limits_{{\cal R}^n}
d^n q\; W(\xi) &=&
|\tilde{\psi}(\underline{p})|^2.  
\label{12}
\end{eqnarray}
 
\noindent
On the other hand if $W(\xi)$ and $W^{\prime}(\xi)$ correspond
respectively to $\hat{\rho}$ and $\hat{\rho}^{\prime}$, it is
easily shown that 
\begin{eqnarray} 
\mbox{Tr}(\hat{\rho}^{\prime}
\hat{\rho}) = (2\pi\hbar)^{-n} \int\limits_{{\cal R}^{2n}}
d^{2n}\xi\; W^{\prime}(\xi) W(\xi)\geq 0.  
\label{13}
\end{eqnarray}

\noindent
But since it is easy to construct cases where the trace on the
left hand side actually vanishes, we can expect that in general
$W(\xi)$ becomes negative in some regions of ${\cal R}^{2n}$.
Indeed, the simplest explicit example showing this is the
expression for the Wigner function for the first excited state
of the harmonic oscillator in one dimension.  Taking $n=1$ and
setting $\hbar=1$ for simplicity, we have 
\begin{eqnarray}
\psi(q) = \displayfrac{\sqrt{2}}{\pi^{1/4}} q\;e^{-
q^2/2}\longrightarrow W(\xi) = \frac{2}{\pi} \left(q^2 + p^2 -
\frac{1}{2}\right)\;e^{-q^2 -p^2}.  
\end{eqnarray}

\noindent
In this context it is interesting to recall the following two
results (again in one dimension) as indicative of the
characteristic features of Wigner distributions:

(i)  \underline{Hudson} \cite{14}  For a pure state $\psi(q)$ the Wigner 
function is pointwise  nonnegative if and only if $\psi(q)$ (and 
hence $W(\xi)$ as well) is a (complex) Gaussian.

(ii) \underline{Folland-Sitaram} \cite{15} If $W(\xi)$ has compact 
support in ${\cal R}^2$, it must vanish identically.

Under the unitary action of $Mp(2n)$ on $\hat{\rho}, W(\xi)$
experiences a simple point transformation: 
\begin{eqnarray}
\hat{\rho}^{\prime}=\overline{U}(S) \hat{\rho}
\overline{U}(S)^{-1} \Longleftrightarrow W^{\prime}(\xi) =
W(S^{-1}\xi), S\;\in\; Sp (2n, R).  
\label{15}
\end{eqnarray}

\noindent
Thus we have covariance under the group $Sp(2n,R)$ which is the
maximal linear homogeneous group mixing $q$'s and $p$'s.  This
combined with the results of eqn.$(\ref{12})$ shows that we recover
the correct marginal probability distributions by integrating
over half the variables in $W(\xi)$ even after action by any
$Sp(2n,R)$ transformation.

The sense in which the definitions $(\ref{10},\ref{11})$ of the Wigner
distribution are dual to the Weyl ordering rule is as follows.
The latter rule associates with each  elementary classical
exponential to a corresponding elementary operator exponential,
\begin{eqnarray} \exp(i\;\underline{\lambda}\cdot \underline{q}
-i\;\underline{\mu} \cdot \underline{p})\longrightarrow
\exp\left(i\;\underline{\lambda} \cdot \hat{\underline{q}} -
i\;\underline{\mu}\cdot\hat{\underline{p}}\right),
\end{eqnarray}

\noindent
where $\underline{\lambda}$ and $\underline{\mu}$ are arbitrary
real vectors in ${\cal R}^n$; and this is then extended by
linearity and Fourier transformation to general classical
functions, say \begin{eqnarray} f(\underline{q},\underline{p})
\equiv f(\xi)\longrightarrow \widehat{F} \end{eqnarray}

\noindent
Then the dual relationship is expressed by the equality of two
ways of computing quantum expectation values: \begin{eqnarray}
\mbox{Tr}\left(\hat{\rho}\exp\left(i\;\underline{\lambda}\cdot
\underline{\hat{q}}-
i\;\underline{\mu}\cdot\underline{\hat{p}}\right)\right) &=&
(2\pi\hbar)^{-n} \int\limits_{{\cal R}^{2n}} d^{2n}\xi
W(\xi)\exp
(i\;\underline{\lambda}\cdot\underline{q}-i\;\underline{\mu}\cdot
\underline{p}),\nonumber\\
\mbox{Tr}\left(\hat{\rho}\widehat{F}\right) &=& (2\pi\hbar)^{-n}
\int\limits_{{\cal R}^{2n}} d^{2n}\xi W(\xi) f(\xi).
\end{eqnarray}

The definition $(\ref{10})$ gives $W(\xi)$ for a given pure state
$\psi(\underline{q})$.  By polarisation we can obtain a
sesquilinear expression: for any two pure states $\psi, \varphi$
we can set up a (generally complex) Wigner distribution
\begin{eqnarray} W_{\psi,\varphi}(\xi) = (2\pi\hbar)^{-n}
\int\limits_{{\cal R}^n}                            d^n
q^{\prime} \psi\left(\underline{q}-\frac{1}{2}
\underline{q}^{\prime} \right)
\varphi\left(\underline{q}+\frac{1}{2}\underline{q}^{\prime}
\right)^* \exp
\left(i\;\underline{p}\cdot\underline{q}^{\prime}/\hbar \right)
, \end{eqnarray}

\noindent
linear in $\psi$ and anti-linear in $\varphi$.  Under complex
conjugation we have \begin{eqnarray} W_{\psi,\varphi}(\xi)^* =
W_{\varphi,\psi} (\xi) , \end{eqnarray}

\noindent
and both the formula $(\ref{13})$ and the $Mp(2n)$ covariance law
$(\ref{15})$ can be easily extended for such objects.  For some
purposes such expressions may be useful, but we do not make much
use of them.

While all of the foregoing is quite familiar, it is useful to
make the following additional remarks.   It is characteristic of
the Heisenberg commutation relations $(\ref{1})$ that even after
quantisation, ie., within quantum mechanics, the possible (sets
of simultaneous) eigenvalues of the (commuting) momenta
$\hat{p}_r$ by themselves do not suffer any quantisation.  Thus
a general set of eigenvalues $p_r, r=1,\ldots, n$ for
$\hat{p}_r$ determines a general point in ${\cal R}^n$, just as
the eigenvalues $q_r$ of the position operators $\hat{q}_r$ do.
It is ultimately this that allows us to describe quantum states
for such systems via Wigner distributions over the classical
phase space $T^* {\cal R}^n \simeq {\cal R}^{2n}$, a general
point $(\underline{q}, \underline{p})$ of which is made up of
(non simultaneous) eigenvalue sets $\underline{q}, \underline{p}$
for the (non commuting) operator sets $\hat{q}_r, \hat{p}_r$.
The appearance and use of the classical phase space here is not
as a result of taking the classical or semiclassical limit of
the quantum theory, but is a way of expressing the exact content
of the quantum theory in a fully c-number language.  The role
and relevance of the groups $Sp(2n,R), Mp(2n)$ in Cartesian
quantum mechanics can really be traced back to these facts; it
makes sense to form canonical linear combinations of Cartesian
$\hat{q}$'s and $\hat{p}$'s.  The importance of these remarks is
seen from a comparison with the case of an angle-angular
momentum pair \cite{13}, and the proper way to set up Wigner distributions
in that case.  We recall this briefly in the next Section, emphasising
the differences compared to the Cartesian situation.

\section{The Wigner distribution in the angle-angular momentum case}

For a classical angle variable $\theta\;\in\;(-\pi,\pi)$,
the configuration space $Q$ is the circle $S^1$; so at the
classical level the phase space or cotangent bundle is the
cylinder $T^* S^1\simeq S^1\times {\cal R}$.  This contains, in
addition to the coordinate $\theta$, a generalised momentum,
$p_{\theta}$ say, which can be any real number:
$p_{\theta}\in\;{\cal R}$.  Now in the quantum situation we
have an angle operator $\widehat{\theta}$ with eigenvalues
$\theta\;\in (-\pi,\pi)$, and a conjugate angular
momentum operator $\widehat{M}$ \underline{whose eigenvalues are
quantised} and are $m=0, \pm 1, \pm 2\ldots,$ ie., $m\;\in
{\cal Z}$ and \underline{not} $m\;\in {\cal R}$.  It is
unnatural in this case to write down a commutation relation
between $\widehat{\theta}$ and $\widehat{M}$; rather their
mutual relationship is best expressed through these eigenvalue
and eigenvector statements: 
\begin{mathletters}\label{21}
\begin{eqnarray} \widehat{\theta}|\theta\rangle &=&
\theta|\theta\rangle , \theta\;\in (-\pi, \pi),\nonumber\\
\langle\theta^{\prime}|\theta\rangle &=& \delta(\theta^{\prime}
-\theta) ; \label{21a}\\ \widehat{M}|m\rangle &=&m\;\hbar|m\rangle,
m\;\in {\cal Z}     ,\nonumber\\ \langle
m^{\prime}|m\rangle &=& \delta_{m^{\prime} m} ;\label{21b}\\
\langle\theta|m\rangle &=& (2\pi)^{-1/2}\cdot \exp
(i\;m\;\theta);\label{21c}\\ \int\limits^{\pi}_{-\pi} d\theta
|\theta\rangle \langle \theta |&=& \sum\limits_{m\in{\cal
Z}} |m\rangle\langle m| = 1. \label{21d}
\end{eqnarray} 
\end{mathletters}

\noindent
The Hilbert space ${\cal H}$ relevant here is $L^2(-\pi, \pi)
\simeq \ell^2$.  Now we define the bounded unitary exponentials
(Weyl exponentials) 
\begin{eqnarray} 
U(n) &=& \exp
(i\;n\;\widehat{\theta}), n\;\in\;{\cal Z}
,\nonumber\\ V(\tau)&=& \exp (-i\;\tau\;\widehat{M}),
\tau\;\in (-\pi, \pi) .  
\label{22}
\end{eqnarray}

\noindent
(We do not need to define the more general $U(\sigma), V(\tau)$
for $\sigma, \tau\; \in\;{\cal R})$.  In contrast to the
Cartesian case where both $\hat{q}$ and $\hat{p}$ are unbounded,
here only $\widehat{M}$ is unbounded.  Then, for a given pure 
state $|\psi\rangle  \in\;{\cal H}$ with wavefunction
$\psi(\theta)=\langle\theta|\psi\rangle$, the Wigner
distribution is a real function $W(\theta,m)$ of \underline{an angle
$\theta$ and an integer $m$} defined as follows:
\begin{eqnarray} 
W(\theta, m) = \frac{1}{2\pi}
\int\limits^{\pi}_{-\pi} d\tau\; \psi (\theta + \tau/2)
\psi(\theta -\tau/2)^* \;e^{-im\tau}, 
\label{23}
\end{eqnarray}

\noindent
the arguments of $\psi$ and $\psi^*$ always being in the range
$(-\pi, \pi)$ via shifts of amounts $\pm 2\pi$.  We note that
the pair $(\theta,m)$ is \underline{not} a point in the
classical phase space $T^* S^1$, just because the ``momentum''
eigenvalue $m$ is quantised.  The definition $(\ref{23})$ reproduces
the marginals correctly: 
\begin{eqnarray}
\int\limits^{\pi}_{-\pi} d\theta~ W(\theta,m)&=& |\langle
m|\psi\rangle|^2 , \nonumber\\ \sum\limits_{m\in{\cal Z}}
W(\theta, m) &=& |\langle \theta|\psi\rangle|^2 .
\label{24}
\end{eqnarray}

\noindent
There is an accompanying dual Weyl operator correspondence as
well: it takes elementary classical exponentials \underline{on
$S^1\times {\cal Z}$} into specific products of the $U$'s and
$V$'s of eqn.$(\ref{22})$: 
\begin{mathletters} 
\begin{eqnarray} \exp
(i\;n\;\theta -i\;\tau\;m) \longrightarrow&& U(n)
V(\tau)e^{-i n \tau/2} = V(\tau) U(n) e^{in\tau/2} ,
\nonumber\\ &&n\;\in\;{\cal Z}, \tau\;\in\;(-\pi, \pi)
; \label{25a}\\ &&\nonumber\\ \langle\psi|U(n) V(\tau)
e^{-in\tau/2}|\psi\rangle &=& \int\limits^{\pi}_{-\pi} d\theta
\sum\limits_{m\in {\cal Z}} W(\theta, m) e^{i(n\theta-\tau
m)} .  
\label{25b}
\end{eqnarray} 
\end{mathletters}

\noindent
(The operator exponentials in $(\ref{25a})$ cannot be combined into
single exponentials).  It is the case that the operators $U(n)
V(\tau) e^{-i n \tau/2}$ for all $n\in{\cal Z}$ and
$\tau\in (-\pi,\pi)$ do form a complete (trace orthonormal)
basis for all operators on ${\cal H}$; and what the Weyl rule
here does is to place any operator $\widehat{F}$ on ${\cal H}$
in correspondence with a classical function $f(\theta, m)$
\underline{on $S^1\times {\cal Z}$, not on $T^* S^1\simeq
S^1\times {\cal R}$}
.

One appreciates that here a certain amount of `quantisation' is
already incorporated into the `classical phase space' structure,
before the Wigner distribution can be defined in a reasonable
way.  There is also no room for the groups $Sp(2,R)$ and $Mp(2)$
.  These characteristic differences compared to the Cartesian
case will get magnified in the case of a general Lie group.

The replacements for Eqns.$(\ref{11},{13})$ of the Cartesian case 
turn out to be as follows: 
\begin{eqnarray}
W(\theta,m)&=&\frac{1}{2\pi}\int_{-\pi}^{\pi} 
<\theta+\tau/2|{\hat \rho}|\theta-\tau/2> e^{-im\tau};\nonumber\\
{\rm Tr}({\hat \rho}^\prime {\hat \rho})&=& \sum_{m\in {\cal Z}}
\int_{-\pi}^{\pi}d\theta~W^\prime (\theta, m)~W(\theta,m).
\label{26}
\end{eqnarray}

Thus from the latter we can see again that in general $W(\theta,m)$, 
though real, can become negative for some arguments. 

In concluding this Section, we mention one interesting case which has no 
Cartesian analogue. Since ${\widehat M}$ has a discrete spectrum, its 
eigenvectors are normalisable, and in that case we find:
\begin{equation}
|\psi> = |m_0>~~~~:~~~~W(\theta, m) = \frac{1}{2\pi}\delta_{m,m_0}.
\end{equation} 
Clearly both of eqns.$(\ref{24})$ are satisfied. 

\section{Classical and Quantum Mechanics on Phase space of a Lie group}

As a preliminary step towards setting up the Wigner distribution
formalism for quantum systems with kinematics based on a general
Lie group, we first briefly recall the important features of the
corresponding classical situation \cite{16}.

Let $G$ be a connected  Lie group of
dimension $n$, and let us regard it as the configuration space
$Q$ of a classical dynamical system.  Then the generalised
coordinate for the system is a variable element $g\in  G$.
The corresponding phase space $T^*Q$ is the cotangent bundle
$T^*G$.  We can describe $T^*G$ in intrinsic purely geometric
terms, which has the advantage of being globally well defined.
However from the point of view of facilitating practical
calculations in any particular case, and so as to avoid being
too cryptic, it is also useful to develop local coordinate based
descriptions of $T^*G$.  We outline the former first, and then
turn to the latter.

\noindent
\underline{Intrinsic descriptions of $T^*G$}

As is well known, every Lie group is a parallelizable
differentiable manifold.  Therefore, if we denote the Lie
algebra of $G$ by $\underline{G}$, and the dual to
$\underline{G}$ by $\underline{G}^*$, it turns out that $T^*G$
is essentially the Cartesian product $G \times \underline{G}^*$.
This equivalence can be established in two equally good ways,
neither of which is preferred.  For definiteness we identify
$\underline{G}$ and $\underline{G}^*$ as the tangent and
cotangent spaces to $G$ at the identity $e$: \begin{eqnarray}
\underline{G} = T_e G , \underline{G}^* = T^*_e  G .
\end{eqnarray}

\noindent
The Lie group $G$ automatically brings with it the set of left
translations $L_g$ and the set of right translations $R_g$.
These are mutually commuting realisations of $G$ by mappings of
$G$ onto itself.  Their definitions and main properties are:
\begin{eqnarray} L_g : g^{\prime}\in
G&\rightarrow & g\;g^{\prime}\in\;G , \nonumber\\ L_{g_{1}}
\circ L_{g_{2}}&=& L_{g_{1}g_{2}} ;\nonumber\\ R_g : g^{\prime}\in G
&\rightarrow & g^{\prime} g^{-1}\in G ,\nonumber \\
R_{g_{1}} \circ R_{g_{2}} &=& R_{g_{1}g_{2}}  ;\nonumber\\ L_{g_{1}}\circ
R_{g_{2}}&=& R_{g_{2}}\circ L_{g_{1}}. \label{27} 
\end{eqnarray}

\noindent
The corresponding tangent maps and pull backs act as nonsingular
linear transformations on the tangent and cotangent spaces
respectively at general points of $G$, according to the
following scheme: 
\begin{mathletters} 
\begin{eqnarray} 
(L_g)_*: T_{g^{\prime}}G &\rightarrow & T_{gg^{\prime}} G ,\nonumber\\
(R_g)_* : T_{g^{\prime}}G &\rightarrow &
T_{g^{\prime}g^{-1}}G;\\ L^*_g : T^*_{g^{\prime}} G &\rightarrow
& T^*_{g^{-1}g^{\prime}}G , \nonumber\\ R^*_g :
T^*_{g^{\prime}}G &\rightarrow & T^*_{g^{\prime}g} G.
\end{eqnarray} 
\end{mathletters}

\noindent
Now introduce dual bases $\{e_r\}, \{e^r\}, r=1,2,\ldots, n$ for
$T_eG, ~T^*_eG$: 
\begin{eqnarray} \underline{G} = T_eG =
\mbox{Sp}\{e_r\}, \underline{G}^*&=& T^*_e G =
\mbox{Sp}\{e^r\},\nonumber\\ \langle e^r, e_s\rangle &=&
\delta^r_s , r, s=1,2,\ldots, n.  
\label{29}
\end{eqnarray}

\noindent
By applying the tangent maps to $\{e_r\}$ at $e$, we obtain two
sets of bases at each $T_gG$, in fact two bases for general
vector fields on $G$: 
\begin{eqnarray} X_r(g) &=& (R_{g^{-1}})_*
(e_r) ,\nonumber\\ \tilde{X}_r(g) &=& (L_{g})_* (-e_r)
;\nonumber\\ T_gG &=& \mbox{Sp}\{X_r(g)\} =
\mbox{Sp}\{\tilde{X}_r(g)\}.  
\label{30}
\end{eqnarray}

\noindent
(The negative sign in the second line is to secure common
commutation relations in eqn.$(\ref{31})$ below).  The vector fields
$\{X_r\}$ are right invariant and are the generators of the left
translations $L_g$, while the vector fields $\{\tilde{X}_r\}$
are left invariant and generate the right translations $R_g$.
Each set obeys the commutation relations (commutators among
vector fields!)  characterising the Lie algebra $\underline{G}$
of $G$ and involving structure constants $f_{rs}^t$:
\begin{eqnarray} 
\protect[X_r, X_s\protect]&=&f_{rs}\;^t \;X_t
,\nonumber\\ \protect[\tilde{X}_r, \tilde{X}_s\protect] &=&
f_{rs}\;^t\; \tilde{X}_t,\nonumber\\ \protect[X_r ,
\tilde{X}_s\protect] &=& 0 .  
\label{31}
\end{eqnarray}

\noindent
We naturally have two dual bases for the cotangent spaces
$T_g^*G$: 
\begin{eqnarray} T^*_gG = \mbox{Sp}\{\theta^r(g)\}&=&
\mbox{Sp}\{\tilde{\theta}^r(g)\} , \nonumber\\
\langle\theta^r(g), X_s(g)\rangle &=&
\langle\tilde{\theta}^r(g), \tilde{X}_s(g)\rangle =
\delta^r_s,\nonumber\\ \theta^r(g) &=& R^*_g(e^r) ,\nonumber\\
\tilde{\theta}^r (g) &=& L_{g^{-1}}^*(-e^r) .  
\label{32}
\end{eqnarray}

\noindent
In terms of these forms, the commutation relations $(\ref{31})$ appear
as the Maurer-Cartan relations:
\begin{eqnarray}
d\theta^r +\frac{1}{2}f_{st}\;^{r}\theta^{s}\wedge \theta^t&=&0,\nonumber\\
d{\tilde \theta}^r +\frac{1}{2}f_{st}\;^{r}{\tilde \theta}^{s}
\wedge {\tilde \theta}^t&=&0.
\end{eqnarray}
  At each $g\;\in \;G$ the
two sets of objects are related by the $n\times n$ matrices
${\cal D}(g)=\left({\cal D}^r_s(g)\right)$ of the adjoint
representation of $G$ (superscript = row, subscript = column
index): \begin{eqnarray} \tilde{X}_r(g) &=& -{\cal D}^s_r(g)
X_s(g) ,\nonumber\\ \theta^r(g) &=& -{\cal D}^r_s(g)
\tilde{\theta}^s(g).  \end{eqnarray}

\noindent
The important point is that all these maps, objects and
relationships are globally well defined.

With this geometric preparation, we can easily see in two ways
why the phase space $T^*G$ is essentially the product $G \times
\underline{G}^*$.  A general `point' in $T^*G$ is a pair
$(g,\omega)$ where $g\in G$ and $\omega \in T_g^*G$.
But we can expand $\omega$  in either of the two bases
$\{\theta^r(g)\}, \{\tilde{\theta}^r(g)\}$ for $T^*_gG$, and use
the expansion coefficients to synthesise elements in
$T_e^*G=\underline{G}^*$: 
\begin{mathletters} 
\begin{eqnarray}
\omega = \omega_r \theta^r(g)&=& R^*_g\left(\omega_r
e^r\right)\;\in T^*_g G\Longleftrightarrow \nonumber\\
&&\omega_0 = \omega_r e^r \;\in \underline{G}^* ;\\
&&\nonumber\\ \omega =-\tilde{\omega}_r\tilde{\theta}^r(g)&=&
L_{g^{-1}}^*\left(\tilde{\omega}_r e^r\right)\; \in T_g^*
G\Longleftrightarrow\nonumber\\ &&\tilde{\omega}_0 =
\tilde{\omega}_r e^r \;\in \underline{G}^* .
\end{eqnarray} 
\end{mathletters}

\noindent
Each of these ways of setting up correspondences gives a
globally well-defined method of identifying $T^*G$ with $G
\times \underline{G}^*$.  For given $\omega\;\in T^*_g G,
\omega_0$ and $\tilde{\omega}_0$ are related by the coadjoint
representation of $G$, since \begin{eqnarray} \tilde{\omega}_r
={\cal D}^s_r(g)\omega_s .  \end{eqnarray}

The above development displays the structure of the classical
phase space $T^*G$ in an intrinsic and globally well defined
manner; in particular it brings out the fact that as a bundle
over the base $G, T^*G$ is trivial.  (In contrast, for example,
$T^*S^ 2$ is nontrivial!).  Now, as stated earlier, we link up
to local coordinate based descriptions more suited to practical
computations and statements of Poisson Bracket relations.

\noindent
\underline{Local coordinate descriptions of $T^*G$}

In general the elements of a Lie group $G$ cannot be described
with the help of coordinates in a globally smooth manner.  In
particular this is so if $G$ is compact.  One has to work with
charts or locally defined coordinates, with well-defined
transition rules in overlaps etc.  For simplicity we will work
within a single chart over some open neighbourhood of the
identity; the setting up of a suitable notation to handle a
collection of charts is in principle quite straightforward but
is omitted.

Let the element $g\;\in\;G$ be labelled by $n$ real
independent continuous coordinates $q^r, r=1,2,\ldots,n$; as a
convention we set $q^r=0$ at $e$.  These $q$'s are numerical
generalised coordinates; especially in case $G$ is compact, each
of them is expected to be an angle type variable.  To the set
of coordinates $q^r$ corresponds the element
$g(q)\;\in\;G$.  We identify the basis elements $e_r, e^r$
for $T_eG$ and $T^*_eG$, eqn.$(\ref{29})$, as 
\begin{eqnarray} 
e_r =\left(\frac{\partial}{\partial q^r}\right)_0 , e^r =(dq^r)_0 .
\label{36}
\end{eqnarray}

\noindent
For practical convenience it is often useful to work with some
faithful matrix representation of $G$.  This has nothing to do
with quantisation per se, but is just a convenient way of
handling $G$ less abstractly than otherwise.  In this sense let
$A(q)$ be some faithful matrix representation of $G$; we
identify its generator matrices and commutation relations by
\begin{eqnarray} 
A(\delta q) &\simeq& 1 - i\;\delta q^r T_r
,\nonumber\\ \protect[T_r, T_s\protect] &=& i\; f_{rs}\;^t\;T_t.  
\label{37}
\end{eqnarray}

\noindent
The product of two elements $A(q^{\prime}), A(q)$ is written as
\begin{eqnarray} 
A(q^{\prime}) A(q) = A(f(q^{\prime};q)) ,
\label{38}
\end{eqnarray}

\noindent
where the $n$ functions $f^r(q^{\prime}; q)$ of $2n$ real
arguments each express the composition law in $G$.  Certain
important auxiliary functions play an important role; their
definitions and some properties are summarised here:

\begin{mathletters} 
\begin{eqnarray} 
\eta^r_s(q) &=&
\left(\displayfrac{\partial f^r}{\partial q^{\prime s}}
(q^{\prime}; q)\right)_{q^{\prime}=0} ,\nonumber\\
\tilde{\eta}^r_s(q) &=& \left(\displayfrac{\partial
f^r}{\partial q^{\prime
s}}(q;q^{\prime})\right)_{q^{\prime}=0} ;\label{39a}\\ &&\nonumber\\
\left(\xi^r_s(q)\right) &=& \left(\eta^r_s(q)\right)^{-1}
,\nonumber\\ \left(\tilde{\xi}^r_s(q)\right) &=&
\left(\tilde{\eta}^r_s(q)\right)^{-1} ;\label{39b}\\ &&\nonumber\\ f(\delta
q; q)&\simeq& q + \eta(q) \delta q,\nonumber\\ f(q; \delta
q)&\simeq& q + \tilde{\eta}(q)\delta q;\label{39c}\\ &&\nonumber\\
\eta^r_s(q) \displayfrac{\partial A(q)}{\partial q^r} &=&-i\;T_s
A(q),\nonumber\\ \tilde{\eta}^r_s(q) \displayfrac{\partial
A(q)}{\partial q^r} &=& -i\;A(q) T_s .  
\label{39d}
\end{eqnarray}
\end{mathletters}

\noindent
(For matrix operations here, superscripts (subscripts) are row
(column) indices).  The  vector fields and one forms in
eqns.(\ref{30}, \ref{32}) have the following local expressions:
\begin{eqnarray} 
X_r =
\eta^s_r(q)\displayfrac{\partial}{\partial q^s}&,& \tilde{X}_r =
- \tilde{\eta}^s_r(q) \displayfrac{\partial}{\partial q^s}
;\nonumber\\ \theta^r = \xi^r_s(q) dq^s &,& \tilde{\theta}^r = -
\tilde{\xi}^r_s (q) dq^s ; 
\label{40}
\end{eqnarray}

\noindent
and the adjoint representation matrices ${\cal D}(g)$ are given
by the product 
\begin{eqnarray} {\cal D}(g(q)) = \xi(q)
\tilde{\eta}(q) .  
\label{41}
\end{eqnarray}

In the sense of classical canonical mechanics when we go to
$T^*G$ we have (local) canonically conjugate momentum variables
$p_r, r=1,2,\ldots,n$; and the basic classical Poisson Bracket
(PB) relations are \begin{eqnarray} \{q^r, p_s\} =
\delta^r_s\;,\; \{q^r, q^s\} = \{p_r, p_s\}=0 .  \end{eqnarray}

\noindent
As for the ranges of these variables, while the structure of $G$
determines the nature of the $q^r$, it is generally assumed that
each $p_r$ ranges independently over the entire real line ${\cal
R}$.  In other words, $T^*_g G\simeq {\cal R}^n$ at each $g\;\in\;G$.

While both the coordinates $q^r$ and the momenta $p_r$ have so
far a local character, it is possible to replace the latter by
certain $q$-dependent linear combinations which are then
globally well-defined.  They express the structure of the phase
space $T ^*G$ in a much more natural way.  We get a clue to
their definitions by noticing, upon combining the PB relations
\begin{eqnarray} \{A(q),p_r\} = \displayfrac{\partial
A(q)}{\partial q^r} \end{eqnarray}

\noindent
with eqn.$(\ref{39d})$, that \begin{eqnarray} \left\{A(q), \eta^r_s(q)
p_r\right\} &=& - i\;T_s A(q) ,\nonumber\\ \left\{A(q),
-\tilde{\eta}^r_s(q) p_r\right\} &=& i\;A(q) T_s .
\end{eqnarray}

\noindent
These relations lead us to define generalised canonical momentum
like variables $J_r, \tilde{J}_r$ as follows: \begin{eqnarray}
J_r = \eta^s_r(q) p_s ,  \tilde{J}_r = -\tilde{\eta}^s_r(q) p_s
.  \end{eqnarray}

\noindent
The connection between the two sets is \begin{eqnarray}
\tilde{J}_r = -{\cal D}^s_r(g(q)) J_s ; \end{eqnarray}

\noindent
and, consistent with eqns. $(\ref{37}, \ref{39d})$, their P.B. relations
are
\begin{eqnarray} 
\{J_r, J_s\} &=& f_{rs}\;^t J_t
,\nonumber\\ \{\tilde{J}_r, \tilde{J}_s\} &=& f_{rs}\;^t
\tilde{J}_t ,\nonumber\\ \{J_r, \tilde{J}_s\} &=& 0 .
\label{47}
\end{eqnarray}

\noindent
The complete coordinate-based description of the basic PB
relations obtaining on $T^*G$ can now be given in many equally
good ways, and we list all of them (allowing for some
repetition):
\begin{mathletters}\label{48} 
\begin{eqnarray} \{q^r, q^s\}
&=& 0 ;\label{48a}\\ &&\nonumber\\ \left\{q^r, J_s\right\} &=& \eta^r_s(q)
,\nonumber\\ \left\{q^r, \tilde{J}_s\right\}&=&
-\tilde{\eta}^r_s(q) ,\nonumber\\ \{A(q), J_r\} &=& -i\; T_r
A(q) ,\nonumber\\ \{A(q), \tilde{J}_r\} &=& i\;A(q) T_r;\label{48b}\\
&&\nonumber\\ \{J_r, J_s\} &=& f_{rs}\;^t J_t ,\nonumber\\
\{J_r, \tilde{J}_s\} &=& 0 ,\nonumber\\ \{\tilde{J_r},
\tilde{J}_s\} &=& f_{rs}\;^t \tilde{J}_t . \label{48c} 
\end{eqnarray}
\end{mathletters}

\noindent
It is thus best to view the set of $J_r$  (or $\tilde{J}_r$) as
the covariant momentum canonically conjugate to the group
element $g\;\in G$ as a generalised coordinate.

At this point, in the present framework, we recognise that the
Lie group underlying the kinematic structure of Cartesian
quantum mechanics for $n$ degrees of freedom, expressed by the
Heisenberg commutation relations $(\ref{1}, \ref{3})$, is the Abelian
translation group $G={\cal R}^n$ in $n$ real dimensions.  In
this case, the coordinates $q^r, r=1,2,\ldots,n$ denoting an
element of ${\cal R}^n$ are globally  well-defined, and
$T^*G=T^*{\cal R}^n\simeq{\cal R}^{2n}$, corresponding to the
Cartesian phase space  $q$ 's and $p$'s.  Due to the group being
Abelian, the structure constants vanish; the $n \times n$
matrices $\eta(q), \xi(q), \tilde{\eta}(q), \tilde{\xi}(q)$ of
eqn.$(\ref{39a}, \ref{39b})$ all reduce to the identity matrix;
the momenta $J_r$ and ${\tilde J}_r$ essentially coincide 
as $J_r =-{\tilde J}_r=p_r$ ;  and the P.B.
relations $(\ref{48})$ reduce to the familiar classical forms for
which the Heisenberg relations $(\ref{1})$ are the quantised version.
We have no difficulty in principle in postulating quantum
kinematics through these commutation relations.

However the angle-angular momentum case briefly described in 
Section III corresponds to the group $G=U(1)\simeq SO(2)$
which is of course also Abelian.  But one immediately sees new
features emerging.  For instance, the angle variable $\theta$ is 
not a globally well-defined coordinate over $G$.  It is also
not very satisfactory, due to operator domain problems, to
postulate simple minded Heisenberg type commutation relations
between $\widehat{\theta}$ and its canonical conjugate
$\widehat{M}$ in the quantum case.  This is over and above the
fact that now $\widehat{M}$ is quantised.  Thus in the $G=SO(2)$
case, it is better to base the treatment on the set of relations
for operators, eigenvalues and eigenvectors collected in
eqns.$(\ref{21})$.

Turning to a general Lie group $G$ where the classical P.B.
structure on $T^*G$ is conveyed by any of the forms given in
eqns.$(\ref{48})$, it should be evident that we should not base the
quantum kinematics on a naive set of commutation relations for
operator forms of the group coordinates $q^r$ and the `momenta'
$J_r, \tilde{J}_r$.  Rather, while the latter can be
satisfactorily handled (and this just involves the
representation theory of $G$), the treatment of the abstract
group element $g$ as a `coordinate operator' after quantisation
has to be handled somewhat differently.

\noindent
\underline{Quantum kinematics for the Lie group case}

We now motivate the forms of the replacements for the Heisenberg
canonical commutation relations (2.1, 2.4) when we consider a
quantum system whose configuration space $Q$ is a Lie group $G$.
Just as we identified $G={\cal R}^n$ for $n$ dimensional Cartesian 
quantum mechanics, where we know that the Schr\"{o}dinger
wave functions are complex valued square integrable functions on
${\cal R}^n$, we should now expect that the Schr\"{o}dinger wave
functions should be complex valued square integrable (in a 
suitable sense) functions on $G$.  The question now is: with what
algebraic operator relations do we replace the earlier canonical
$\hat{q}-\hat{p}$  commutation relations?

If we try to avoid the use of (local) coordinates for group
elements, in the interests of being as intrinsic as possible, we
might be tempted to imagine the following: upon quantisation,
the classical generalised coordinate $g\in G$ is replaced
by an ``operator $\hat{g}$~'' for which the possible
``eigenvalues'' are the classical abstract group elements!
However this seems excessively formal.  A more reasonable
strategy would be to first set up a classical commutative
algebra ${\cal A}$, say, of all smooth, i.e., ${\cal C}^\infty$,  
real valued functions $f(g)$ on $G$: 
\begin{eqnarray} 
g\;\in \;G
&\rightarrow & f(g)\;\in\;{\cal R} : f\;\in\;
{\cal A};\nonumber\\ f_1, f_2 \;\in\;{\cal A} &\Rightarrow
& c_1 f_1 + c_2 f_2, f_1 f_2 \;\in\;{\cal A} .
\end{eqnarray}

\noindent
Here the $c$'s are real numbers, and the above  choice of
functions $f\;\in {\cal A}$ captures the differentiable manifold structure 
of $G$.  The left
and right translations $L_g, R_g$ of eqn.$(\ref{27})$ now act on ${\cal
A}$ as follows: 
\begin{eqnarray} \mbox{Left action}&:&
g^{\prime}\in G : f(g)\rightarrow f(g^{\prime -1} g)
;\nonumber\\ \mbox{Right action} &:& g^{\prime}\in \; G :
f(g)\rightarrow f(g g^{\prime}) .  
\label{50}
\end{eqnarray}

\noindent
Upon quantisation we ask for an Abelian operator algebra
$\widehat{{\cal A}}$, say, consisting of hermitian operators
such that in a natural way we ensure 
\begin{eqnarray}
f\;\in {\cal A}&\rightarrow&
\hat{f}\;\in \;\widehat{{\cal A}} ; \nonumber\\ f_1,
f_2\;\in  {\cal A}&\Rightarrow& c_1f_1 + c_2 f_2\rightarrow
c_1 \hat{f}_1 + c_2 \hat{f}_2,\nonumber\\ &&f_1 f_2 \rightarrow
\hat{f}_1 \hat{f}_2 .  
\label{51}
\end{eqnarray}

\noindent
This is the replacement for the $\hat{q}-\hat{q}$ part of the
canonical relations $(\ref{1})$, and is the quantised version of the
PB relations $\{q^r, q^s\}=0$ in eqn.$(\ref{48a})$, in a globally
well-defined form.

Turning to the quantisation of the remaining PB relations in
eqns.$(\ref{48b},\ref{48c})$, we can work either with finite group elements or
with infinitesimal generators.  In the former, we ask for
unitary operator families $V(g), \tilde{V}(g)$ realising the
left and right translation groups on $G$, and producing on
$\widehat{{\cal A}}$ the effects implied by eqn. $(\ref{50})$:

\begin{mathletters} 
\begin{eqnarray} f(g)\;\in \;{\cal
A}&\rightarrow& \hat{f}\; \in \widehat{{\cal A}} \Rightarrow
\nonumber\\ f(g^{\prime -1}g)&\rightarrow& V(g^{\prime}) \hat{f}
V(g^{\prime})^{-1} , \nonumber\\ f(g g^{\prime})&\rightarrow&
\tilde{V}(g^{\prime}) \hat{f} \tilde{V} (g^{\prime})^{-1},
g^{\prime}\;\in G ;\label{52a}\\ &&\nonumber\\ V(g_1)V(g_2)&=& V(g_1
g_2) ,\nonumber\\ \tilde{V}(g_1) \tilde{V}(g_2)&=& \tilde{V}
(g_1g_2) ;\label{52b}\\ &&\nonumber\\ V(g_1) \tilde{V}(g_2) &=&
\tilde{V}(g_2) V(g_1) .
\label{52c}  
\end{eqnarray} 
\end{mathletters}

\noindent
The operator relations $(\ref{52a})$ are the quantised and finite
forms of the PB relations in $(\ref{48b})$ involving
$\left\{q^r\;\mbox{or}\;A(q), J_s\;\mbox{or}\;
\tilde{J}_s\right\}$; while the operator relations $(\ref{52b}, \ref{52c})$
are the integrated forms of the result of quantising the PB
relations $(\ref{48c})$ keeping track of course of the global connectivity 
properties of $G$.  The latter can also be expressed at the generator
level.  If the generators of $V(g), \tilde{V}(g)$ are
$\widehat{J}_r, \widehat{\tilde{J}}_r$ respectively, we require
them to be hermitian and to obey 
\begin{eqnarray}
\protect\left[\widehat{J}_r, \widehat{J}_s\protect\right] &=&
i\;f_{rs}\;^t \widehat{J}_t ,\nonumber\\
\protect\left[\widehat{\tilde{J}}_r,
\widehat{\tilde{J}}_s\protect \right] &=& i\;f_{rs}\;^t
\widehat{\tilde{J}}_t ,\nonumber\\ \protect\left[\widehat{J}_r,
\widehat{\tilde{J}}_s\protect\right]&=& 0, \nonumber\\
\widehat{\tilde{J}}_r =& -&{\cal D}_{r}^{s}(g) {\hat J}_s.
\label{53}
\end{eqnarray}

In comparison to the canonical commutation relations $(\ref{1})$, we
see that eqn.$(\ref{52a})$ correspond to the $\hat{q}-\hat{p}$ part,
and eqns. $(\ref{52b}, \ref{52c}, \ref{53})$ correspond to the 
$\hat{p}-\hat{p}$ part, respectively.  Thus the complete set of algebraic
relations expressing quantum kinematics for quantum mechanics on
a Lie group as configuration space are eqns. $(\ref{51}, \ref{52a}-\ref{52c}, 
\ref{53})$.
These have to be realised irreducibly on a suitable Hilbert
space.

A clarifying remark may be made at this point.  If we were
looking only for a unitary representation (UR) or unitary
irreducible representation (UIR) of $G$, the only commutation
relations to be satisfied would be those among the hermitian
generators, $\widehat{J}_r$ say, of such a UR or UIR.  But
these comprise only a part - the $\hat{p}-\hat{p}$ part - of the
complete set of algebraic relations developed above; and do not
include the operators in $\widehat{\cal A}$ which represent smooth
functions on $G$ and which capture the notion of position
operator in this case.  Conversely, a single UIR of $G$ on some
Hilbert space, over which the $\widehat{J}_r$ act irreducibly,
is here the analogue of a single simultaneous (ideal)
eigenvector of all the (commuting) momenta $\hat{p}_r$.  The
latter is always one dimensional because ${\cal R}^n$ is Abelian
- there is just one (ideal) eigenvector $|\underline{p}\rangle$
of the $\hat{p}_r$ for given eigenvalues $p_r$.  With a general
non Abelian Lie group $G$, the analogue of a ``momentum
eigenstate'' is a (finite or infinite dimensional) UIR of $G$.

A natural representation of all the algebraic relations imposed
above is via the regular representation of $G$.  We will
hereafter always assume that there is a unique (upto a factor)
left and right translation invariant volume element $dg$ on $G$,
the Haar measure, which in the compact case will be normalised
so that $G$ has total volume unity: \begin{eqnarray} f\;\in 
{\cal A}: \int\limits_G dg\;f(g) &=& \int\limits_G dg\;
f(g^{\prime -1}g) = \int\limits_G dg\;f(g g^{\prime}) ;
\nonumber\\ \int\limits_G dg &=& 1\;\mbox{if}\;G\;\mbox{compact}
\end{eqnarray}
In local coordinates $q^r$ for $G$, apart from a normalisation factor, this 
volume element involves the determinants of the matrices $\xi(q), 
~{\tilde \xi}(q)$ defined in $(\ref{39b})$ :
\begin{equation}
dg= {\rm det}(\xi(q))d^n q ={\rm det}({\tilde \xi}(q))d^n q 
\end{equation}
\noindent
Then the Hilbert space ${\cal H} = L^2(G)$  is defined, in the
``Schr\"{o}dinger representation'', as 
\begin{eqnarray} {\cal H}
= \left\{\psi(g) \in  {\cal C}| \parallel\psi\parallel^2 =
\int\limits_G dg |\psi(g)|^2 < \infty\right\} 
\label{55}
\end{eqnarray}

\noindent
On this space the required operators $\hat{f}\;\in
\widehat{{\cal A}}, V(g^{\prime}), \tilde{V}(g^{\prime})$ 
are easily defined  : 
\begin{eqnarray}
f(g)\;\in {\cal A}\rightarrow \hat{f}\in
\widehat{{\cal A}} &:& (\hat{f}\psi)(g) = f(g) \psi(g)
;\nonumber\\ (V(g^{\prime})\psi)(g) &=& \psi(g^{\prime
-1}g),\nonumber\\ (\tilde{V}(g^{\prime})\psi)(g) &=& \psi(g
g^{\prime}) .  
\label{56}
\end{eqnarray}

\noindent
This is indeed an irreducible representation of the complete
algebraic system as is shown in Appendix A.

In local coordinates if we write $\psi(g)$ as $\psi(q)$, the
generators $\widehat{J}_r, \widehat{\tilde{J}}_r$ are
immediately obtained as 
\begin{eqnarray} 
\widehat{J}_r &=&
-i\;\eta^s_r(q)\frac{\partial}{\partial q^s}, \nonumber\\
\widehat{\tilde{J}}_r &=& i\;\tilde{\eta}^s_r(q)\frac{\partial}
{\partial q^s} ,\nonumber\\ V(\delta q)&\simeq& 1 -i\;\delta q^r
\widehat{J}_r, \tilde{V}(\delta q) \simeq 1 - i\;\delta q^r
\widehat{\tilde{J}}_r .  
\label{57}
\end{eqnarray}

\noindent
Thus these generators are essentially the vector fields $X_r,
\tilde{X}_r$ defined earlier in eqns. $(\ref{30}, \ref{40})$, but now
interpreted as hermitian operators  on $L^2(G)$.

It is for the elements $|\psi\rangle$ in the Hilbert space
${\cal H}$ of eqn.$(\ref{55})$ that we wish to set up a Wigner
distribution formalism with natural properties.
     
\section{New features to be accommodated}

For Cartesian quantum mechanics we have the well-known Stone-von
Neumann theorem which states that upto unitary equivalence there
is only one irreducible representation of hermitian operators
$\hat{q}_r, \hat{p}_r$ obeying the Heisenberg relations $(\ref{1})$.
This irreducible representation is of course describable in many
ways - position representation with $\hat{q}_r$ diagonal;
momentum representation with $\hat{p}_r$ diagonal; Fock basis;
coherent states etc.. When ${\cal R}^n$ here is replaced by a general
Lie group $G$, we have already appreciated that the basic
building block for the quantum theory is \underline{not} a UIR
of $G$, but an irreducible representation of the entire
algebraic system consisting of $\widehat{{\cal A}}, 
V (\cdot)$ and $\tilde{V}(\cdot)$. A single UIR of $G$
is too small to support the action of a group element as a
generalised coordinate.  To achieve this many UIR's of $G$ have
to be put together in a careful manner.

As for UIR's of $G$, we recall several familiar facts.  If $G$
is compact, every UIR is finite dimensional.  If $G$ is
non compact simple, then every nontrivial UIR is infinite
dimensional; every finite dimensional representation is
non unitary; and in addition there are infinite dimensional
non unitary representations.

We will for the most part and for definiteness consider the case
of a compact simple Lie group $G$.  A natural irreducible
representation of the entire algebraic structure we are
interested in is given, as we have seen, by the regular
representation.  The main features and auxiliary operators
associated with it, and some notations, are given in Appendix A.

In passing we mention the fact that while for compact $G$ every
UIR is `seen' in the regular representation, in the non compact
case there are UIR's (the exceptional series) not contained in
the regular representation. 

 One last general comment is important before proceeding.  As we
have seen in general the `momenta' of our problems are
non commuting operators.  This is a genuine new feature absent in
Cartesian quantum mechanics, and it has significant
consequences.  We have seen hints of this in the angle-angular
momentum case in Section III, even though there was only one
momentum $\widehat{M}$ involved.  For general $G$, the momenta
$\widehat{J}_r, \widehat{\tilde{J}}_r$ cannot all be
simultaneously diagonal and their spectra undergo quantisation.
Therefore the space of arguments of the Wigner distribution has
to be carefully chosen; it is definitely not a function on
the classical phase space $T^{*}G$ in general.  By the same token, there are 
in general no analogues to the groups $S p(2n,R), Mp(2n)$ which 
are so important in the Cartesian case.

\section{The Wigner distribution in the Regular Representation}

Let $|\psi\rangle\in{\cal H}=L^2(G)$ be a normalised state
vector.  The corresponding ``position space'' probability
density is a probability distribution on the group $G$ given by
(cf.eqn.$(\ref{A.5}))$.  
\begin{eqnarray} |\psi(g)|^2 = |\langle
g|\psi\rangle|^2 .  
\label{58}
\end{eqnarray}

\noindent
The complementary ``momentum space'' probability distribution is
(assuming $G$ to be compact) a discrete set of probabilities
indexed by the quantum numbers $JMN$ and given by
(cf.eqn.$(\ref{A.12})$): 
\begin{eqnarray} |\psi_{JMN}|^2 = |\langle
JMN|\psi\rangle|^2 .  
\label{59}
\end{eqnarray}

\noindent
The common normalisation states that \begin{eqnarray} \parallel
\psi \parallel^2 = \int\limits_G dg |\psi(g)|^2 = \sum\limits
_{JMN} |\psi_{JMN}|^2 = 1 .  \end{eqnarray}

At first glance we might suppose that, given $|\psi\rangle$, the
corresponding Wigner distribution $W(\ldots)$ should be a real
function with $g$ and $JMN$ (coordinates and quantised momenta)
as arguments, bilinear in $\psi$ (more precisely involving one
$\psi$ factor and one $\psi^*$ factor), such that integration
over $g$ yields $|\psi_{JMN}|^2$ while  summation over $JMN$
yields $|\psi(g)|^2$.  This would be a natural way in which the
marginals $(\ref{58}, \ref{59})$ are reproduced.  However we should also
require covariance under both (left and right) actions by $G$
on $\psi$ : the choice of the arguments in $W(\ldots)$ should
allow for a natural linear transformation law under each of the
changes $\psi(g)\rightarrow\psi(g^{-1}_1 g)$ and
$\psi(g)\rightarrow \psi(gg_2)$ in $\psi$.  Now the `momentum space'
amplitudes $\psi_{JMN}$ of $\psi$ transform linearly as follows
(cf. eqn.$(\ref{A.12})$).  
 
\begin{eqnarray}
|\psi^{\prime}\rangle &=& V(g_1)|\psi\rangle , \psi^{\prime}(g)
= \psi\left(g^{-1}_1 g\right) :\nonumber\\ \psi^{\prime}_{JMN}
&=& \sum\limits_{M^{\prime}} {\cal D}^J_{MM^{\prime}} (g_1)^*
\psi_{JM^{\prime}N} ;\nonumber\\|\psi^{\prime\prime}\rangle &=&
\tilde{V}(g_2)|\psi\rangle , \psi^{\prime\prime}(g) = \psi(gg_2)
:\nonumber\\ \psi^{\prime\prime}_{JMN} &=&
\sum\limits_{N^{\prime}} {\cal D}^J_{N^{\prime}N}
\left(g^{-1}_2\right)^* \psi_{JMN^{\prime}}.  
\label{61}
\end{eqnarray}

\noindent
Thus in each case there is a linear mixing of the components
$\psi_{JMN}$ for fixed $J$ at the $\psi$ level.  Remembering
that $W(\ldots)$ should involve the bilinear expressions
$\psi\psi^*$, a little reflection shows that it would be too
narrow to imagine that the Wigner distribution should be some
real function $W(g;JMN)$: there would be too few momentum space
arguments to support the changes $(\ref{61})$ in $\psi$ in a reasonable
manner.

There is another way in which this situation could be described.
As we have already pointed out  in Section V, an essential new
feature is that now the analogue of the single momentum eigenket
$|\underline{p}\rangle$ in Cartesian quantum mechanics is a
multidimensional  object, an entire UIR of $G$; actually in the
regular representation even more since both $\widehat{J}$'s and
$\widehat{\tilde{J}}$'s have to be represented.  In this sense,
with a general Lie group $G$ different from ${\cal R}^n$, there
is a genuine asymmetry between positions and momenta.  While the
analogue of `position eigenstate' remains one dimensional,
$|\underline{q}\rangle$ being replaced by $|g\rangle$, the
momentum operators constitute the non commutative algebra of
$\widehat{J}$'s and $\widehat{\tilde{J}}$'s, leading to the
quantum numbers $JMN$ where only $J$ remains fixed.
(Incidentally the first part of this statement is not in
conflict with the fact that $G$ itself may be non Abelian!  In
local coordinates  $q^r$ for $G$, the ideal ket $|g\rangle$ may
be written as $|\underline{q}\rangle$, and all the $q$'s are
simultaneously diagonal).  Out of all the momentum operators, a
complete commuting set consists of the (shared) Casimir
operators formed out of the $\widehat{J}$'s, and separately out
of the $\widehat{\tilde{J}}$'s, accounting for $J$ in the set
$JMN$; a maximal commuting subset of the $\widehat{J}$'s,
supplying some of the labels in $M$; a similar maximal commuting
subset of the $\widehat{\tilde{J}}$'s supplying the analogous
labels in $N$; and further nonlinear mutually commuting
expressions in $\widehat{J}$'s (respectively
$\widehat{\tilde{J}}$'s) to account for the remaining labels in
$M$ (respectively $N$).  The  main point is that in the process
of obtaining the marginal distribution $(\ref{59})$ upon integrating
the Wigner distribution with respect to its argument $g$, we
should expect at first to get something like a density matrix
within the $J$th subspace of momentum labels, and then upon
going to the diagonal elements recover the probabilities
$|\psi_{JMN}|^2$.  The transformation laws $(\ref{61})$ can already be
written in a matrix form (at the level of $\psi$, not of the
density matrix) thus: \begin{eqnarray} \psi^{(J)} = (\psi_{JMN})
: |\psi\rangle&\rightarrow & V(g_1)
\tilde{V}(g_2)|\psi\rangle\Rightarrow \nonumber\\ \psi^{(J)}
&\rightarrow & {\cal D}^J(g_1)^* \psi^{(J)} {\cal D}^J \left(g^{-1}_2
\right)^* .  \end{eqnarray}

\noindent
Since the individual probabilities $|\psi_{JMN}|^2$ do not
transform linearly among themselves under such $G$ actions, but
do bring in `off-diagonal' quantities, the structure of the
Wigner distribution will inevitably reflect this fact.

Based on these considerations we now list the basic desired
properties for the Wigner distribution $W(\ldots)$ associated
with a given normalised $|\psi\rangle \in  {\cal H}$ (for
simplicity the dependence of the former on the latter is left
implicit) initially as: 
\begin{mathletters}\label{63} 
\begin{eqnarray}
\psi(g)\; \in  {\cal H}\rightarrow W(g;
JMN\;M^{\prime}N^{\prime}) ,&&\nonumber\\ W(g;
JMN\;M^{\prime}N^{\prime})^* &=&W(g;JM^{\prime}N^{\prime}\;MN)
;\label{63a}\\ \int\limits_G dg \;W(g; JMN\;MN) &=& |\psi_{JMN}|^2 ,
\nonumber\\ \sum\limits_{JMN} W(g; JMN\;MN) &=& |\psi(g)|^2 ;\label{63b}\\
\psi^{\prime}(g) =\psi(g^{-1}_1 g)\rightarrow &&\nonumber\\
W^{\prime}(g;JMN\;M^{\prime}N^{\prime})
&=&\sum\limits_{M_{1}M_{1} ^{\prime}} {\cal D}^J_{MM_{1}}(g_1)
D^J_{M^{\prime}M^{\prime}_{1}} (g_1)^* W\left(g^{-1}_1 g;
JM_{1}N\;M^{\prime}_1 N^{\prime}\right), \label{63c}\\
\psi^{\prime\prime}(g) = \psi(g g_2)\rightarrow &&\nonumber\\
W^{\prime\prime}(g; JMN\;M^{\prime}N^{\prime})&=&
\sum\limits_{N_1 N_1^{\prime}} W\left(g g_2; JMN_1 M^{\prime}
N^{\prime}_1\right) {\cal D}^J_{N_1 N}\left(g_2^{-1}\right)
{\cal D}^J_{N^{\prime}_1 N^{\prime}}\left(g^{-1}_2\right)^*.
\label{63d}
\end{eqnarray} 
\end{mathletters}

\noindent
One can see that the covariance conditions $(\ref{63c},\ref{63d})$ are 
compatible with the transformation laws $(\ref{61})$ for $\psi_{JMN}$ 
and the requirement $(\ref{63b})$ for reproduction of the marginals. Actually 
one has little option but to extend the requirement in the first 
of eqn. $(\ref{63b})$ to read: 
\begin{eqnarray} \int\limits_{G} dg
W(g; JMN\;M^{\prime}N^{\prime}) =
\psi^*_{JMN}\;\psi_{JM^{\prime}N^{\prime}}.  
\label{64}
\end{eqnarray}

\noindent
Upon then setting $M^{\prime}=M, N^{\prime}=N$ here one recovers
the true probabilities $|\psi_{JMN}|^2$.  To all of the above we
add a natural condition that $W$ be of the general structure
$\psi\psi^*$.

We now propose the following form for the Wigner distribution:
\begin{eqnarray} W(g; JMN\;M^{\prime}N^{\prime}) &=&
N_J\int\limits_G dg^{\prime} \int\limits_G dg^{\prime\prime}\;
\delta\left( g^{-1}
s(g^{\prime},g^{\prime\prime})\right)\nonumber\\ &&{\cal D}^J_{MN}
(g^{\prime})\;\psi(g^{\prime})^*\;
{\cal D}^J_{M^{\prime}N^{\prime}}(g^{\prime\prime})^*\;
\psi(g^{\prime\prime}).  
\label{65}
\end{eqnarray}

\noindent
This involves a group element
$s(g^{\prime},g^{\prime\prime})\;\in \;G$ depending on two
arguments also drawn from $G$, which must have suitable
covariance and other properties.  The set of conditions $(\ref{63a}-
\ref{63d},\ref{64})$
now translates into a set of requirements on this function $s:
G\times G\rightarrow G$ which are: 
\begin{eqnarray} g^{\prime},
g^{\prime\prime}\;\in \;G &\rightarrow&
s(g^{\prime},g^{\prime\prime})\;\in \; G ;\nonumber\\
s(g^{\prime}, g^{\prime\prime})&=& s(g^{\prime\prime},
g^{\prime}) ;\nonumber\\ s(g^{\prime}, g^{\prime}) &=&
g^{\prime} ;\nonumber\\ s\left(g_1 g^{\prime} g^{-1}_2 , g_1
g^{\prime\prime} g^{-1}_2\right) &=& g_1 \;s(g^{\prime},
g^{\prime\prime}) g_2^{-1} .  
\label{66}
\end{eqnarray}

\noindent
{\it Any choice of a function $s(g^{\prime}, g^{\prime\prime})$
obeying these conditions leads to an acceptable definition of a
Wigner distribution for quantum mechanics on a (compact) Lie
group $G$}

The second and third lines of eqn.$(\ref{66})$ suggest that we view
$s(g^{\prime}, g^{\prime\prime})$ as a kind of symmetric square
root of the product of two (generally non commuting) group
elements $g^{\prime}, g^{\prime\prime}\;\in \;G$.  The
covariance conditions in the  last line help us simplify the
problem to the choice of a suitable function $s_0(g^{\prime})$
of a single argument drawn from $G$, obeying conditions that
ensure $(\ref{66})$: 
\begin{eqnarray} s(e,g) &=& s_0(g) ,\nonumber\\
s(g^{\prime},g^{\prime\prime})&=& g^{\prime} \;s_0(g^{\prime -
1}g^{\prime\prime}) :\nonumber\\ s_0(e) &=& e ;\nonumber\\
s_0(g^{-1}) &=& g^{-1} s_0(g) ;\nonumber\\ s_0(g^{\prime} g
g^{\prime -1}) &=& g^{\prime}\;s_0(g) g^{\prime -1} .
\label{67}
\end{eqnarray}

\noindent
It is a consequence of these conditions on $s_0(g)$ that
\begin{eqnarray} 
s_0\;(g)\; g = g\;s_0\;(g) .  
\end{eqnarray}

We now present a solution to the above problem in the case of a
compact simple Lie group $G$.  Any such group carries a unique
Riemannian metric defined in terms of the structure constants,
and possessing left and right translation invariances.  We shall
content ourselves with a local coordinate description and use
the notations of eqn.$(\ref{36} - \ref{41})$.  Admitting the over use 
of the letter $g$, at the identity the metric tensor has components
\begin{eqnarray} 
g_{rs} (0)= - f_{ru}\;^v\; f_{sv}\;^u ,
\label{69}
\end{eqnarray}

\noindent
the negative sign ensuring that the matrix $(g_{rs}(0))$ is
positive definite.  This tensor is checked to be invariant under
the action by the adjoint representation: 
\begin{eqnarray} 
{\cal D}^u_r(g) {\cal D}^v_s(g)  g_{uv}(0) = g_{rs}(0) .
\label{70}
\end{eqnarray}

\noindent
At a general point $g(q)\;\in \;G$ we obtain $g_{rs}(q)$ by
shifting $g_{rs}(0)$ as a tensor to $g(q)$ using either left or
right translation; on account of $(\ref{70})$ the two results are the
same and we find: 
\begin{eqnarray} 
g_{rs}(q) &=& \xi^u\;_r(q)
\xi^v\;_s(q) g_{uv}(0) \nonumber\\ &=& \tilde{\xi}^u\;_r(q)
\tilde{\xi}^v\;_s(q) g_{uv}(0) .  
\label{71}
\end{eqnarray}

Geodesics in $G$ are curves of minimum length with respect to
the above Riemannian metric.  As is well known, both left and
right translations, $L_g$ and $R_g$, applied pointwise map
geodesics onto geodesics.  Thus if $g(q(\sigma))\;\in \;G$
is a solution to the variational problem 
\begin{eqnarray}
\delta \int\limits^{\sigma_{2}}_{\sigma_{1}} d\sigma
\left(g_{rs}(q(\sigma)) \frac{dq^r(\sigma)}{d\sigma}
\frac{dq^s(\sigma)}{d\sigma}\right)^{1/2} = 0 , 
\label{72}
\end{eqnarray}

\noindent
where we assume an affine parametrisation is chosen so that
\begin{eqnarray} g_{rs}(q(\sigma))\frac{dq^r(\sigma)}{d\sigma}
\frac{dq^s(\sigma)}{d\sigma} =\mbox{constant} , \end{eqnarray}

\noindent
then both $L_{g_{1}} g(q(\sigma))$ and $R_{g_{2}} g(q(\sigma))$
are solutions to the same  variational problem.

We now use geodesics in $G$ to construct the function $s_0(g)$.
It is a fact that for almost all $g\;\in \;G$ (ie., except
for a set of measure zero), there is a unique geodesic
(minimising the functional appearing in eqn.$(\ref{72})$) running from
the identity $e$ to $g$.  We assume the affine parametrisation
is normalised so that the geodesic passes through $e$ at
$\sigma=0$ and through $g$ at $\sigma=1$: \begin{eqnarray}
g\;\in \;G : g(q(0)) = e,\;g(q(1))=g .  \end{eqnarray}

\noindent
We then take $s_0(g)$ to be the half way point reached at
$\sigma=1/2$: \begin{eqnarray} s_0(g) = g(q(1/2)) .
\label{75}
\end{eqnarray}

\noindent
It is a matter of easy verification that all the conditions
$(\ref{67})$ are indeed obeyed: one has to exploit the natural
covariance and other properties of general geodesics.  With this
we have solved the problem of defining Wigner functions for
quantum mechanics on a (compact) Lie group, possessing all the
properties listed in eqn.$(\ref{63a}-\ref{63d},\ref{64})$.  The fact that 
$s_0(g)$ is
defined everywhere except possibly on a set of vanishing measure
causes no problems in carrying out integrations over $G$, or in
recovering the marginals.

It may be pointed out that for a general pair of elements
$g^{\prime}, g^{\prime\prime}\;\in \;G$ (except in cases
amounting to a set of vanishing measure) there is a unique
geodesic running from $g^{\prime}$ to $g^{\prime\prime}$,
normalised so that the affine parameter has values $\sigma=0$
and $\sigma=1$ at start and at finish.  This geodesic is the
result of applying $L_{g^{\prime}}$ to the geodesic from $e$ to
$g^{\prime -1} g^{\prime\prime}$, or equally well of applying
$R_{g^{\prime -1}}$ to the one from $e$ to $g^{\prime\prime}
g^{\prime -1}$.  In either view,
$s(g^{\prime},g^{\prime\prime})$ is the mid point of this
geodesic, reached at $\sigma=1/2$.  Moreover, geodesics passing
through the identity $e$ are one-parameter subgroups in $G$.  If
we define $s_0(g)$ in eqn.$(\ref{75})$ to be the square root of the
element $g$, we can write the general quantity $s(g^{\prime},
g^{\prime\prime})$ in these suggestive ways: \begin{eqnarray}
s(g^{\prime},g^{\prime\prime}) &=& g^{\prime} (g^{\prime -1}
g^{\prime\prime})^{1/2} = 
g^{\prime\prime}(g^{\prime\prime -1}
g^{\prime})^{1/2}\nonumber\\ &=& (g^{\prime\prime}g^{\prime
-1})^{1/2} g^{\prime}=(g^{\prime}g^{\prime\prime -1})^{1/2}
g^{\prime\prime}.  \end{eqnarray}

The definition $(\ref{65})$ of the Wigner distribution associated with a pure
state $\psi(g)$ generalises to a mixed state with density operator 
${\hat \rho}$:
\begin{eqnarray} W(g; JMN\;M^{\prime}N^{\prime}) &=&
N_J\int\limits_G dg^{\prime} \int\limits_G dg^{\prime\prime}\;
\delta\left( g^{-1}
s(g^{\prime},g^{\prime\prime})\right) 
<g^{\prime\prime}|{\hat \rho}|g^{\prime}>\nonumber\\ &&~~~~~~~~~~~~~~
{\cal D}^J_{MN}
(g^{\prime})\;
{\cal D}^J_{M^{\prime}N^{\prime}}(g^{\prime\prime})^*,\;\nonumber\\
\int dg~ W(g; JMN\;M^{\prime}N^{\prime})&=& <JM^{\prime}N^{\prime}|
{\hat \rho}|JMN>, \nonumber\\
\sum_{JMN} W(g; JMN\;MN)&=&<g|{\hat \rho}|g>  
\label{77}
\end{eqnarray}
We now verify that $W(g; JMN\;M^{\prime}N^{\prime})$ is a faithful
representation of ${\hat \rho}$ in the sense that it contains complete
information concerning ${\hat \rho}$. This will be shown by developing
analogues to the previous eqns. $(\ref{13}, \ref{26})$; in fact we will 
find two separate analogues.

The Wigner function in eq.$(\ref{77})$ transforms according to eqns.
$(\ref{63c},\ref{63d})$ under independent left and right translations. By
setting $N=N^\prime$ and then summing over $N$, we obtain, 
using $(\ref{A.9a})$, a slightly simpler function , ${\tilde W}$ say, 
corresponding to the density operator ${\hat \rho}$ : 
\begin{eqnarray}
{\tilde W}(g; JMM^\prime)&=&\sum_{N} W(g; JMN\;M^{\prime}N)\nonumber\\
&=&N_J\int\limits_G dg^{\prime} \int\limits_G dg^{\prime\prime}\;
\delta\left( g^{-1}
s(g^{\prime},g^{\prime\prime})\right) 
<g^{\prime\prime}|{\hat \rho}|g^{\prime}>
{\cal D}^J_{MM^\prime}
(g^{\prime}{g^{\prime\prime}}^{-1})\;.
\label{78}
\end{eqnarray}
This auxiliary function is invariant under right translations except for a
change of argument $g\rightarrow gg_2$, while under left translations it
transforms in a manner similar to eq.$(\ref{63d})$. 
Now consider two density operators ${\hat \rho}_1$ and ${\hat \rho}_2$, with
associated functions ${\tilde W}_1$ and ${\tilde W}_2$. It  can then be shown 
that we can obtain ${\rm Tr}({\hat \rho}_1{\hat \rho}_2)$ from 
${\tilde W}_1$ and ${\tilde W}_2$ by `summing' over all the arguments: 
\begin{equation}
\sum_{JMM^\prime} N_{J}^{-1}\int dg {\tilde W}_1(g; JMM^\prime)
{\tilde W}_2(g; JM^\prime M)={\rm Tr}({\hat \rho}_1{\hat \rho}_2). 
\label{78a}
\end{equation}
The proof is presented in Appendix B. Since any density operator 
${\hat \rho}_1$ is fully determined by the traces of its products  with all 
other density operators ${\hat \rho}_2$, we can see that even the simpler 
function ${\tilde W}(g; JMM^\prime)$ fully characterises ${\hat \rho}$. 

Obviously another analogue to eqns.$(\ref{13}, \ref{26})$ can be obtained by
interchanging the roles of left and right translations in the above. If in 
place of $(\ref{78}) $ we define 
\begin{eqnarray}
{\tilde{\tilde W}}(g; JNN^\prime)&=&\sum_{M} W(g; JMN\;MN^\prime)\nonumber\\
&=&N_J\int\limits_G dg^{\prime} \int\limits_G dg^{\prime\prime}\;
\delta\left( g^{-1}
s(g^{\prime},g^{\prime\prime})\right) 
<g^{\prime\prime}|{\hat \rho}|g^{\prime}>
{\cal D}^J_{N^\prime N}
({g^{\prime\prime}}^{-1}{g^{\prime}})\;,
\label{80}
\end{eqnarray}
then for two density operators ${\hat \rho}_1,~{\hat \rho}_2$ we have 
\begin{equation}
\sum_{JNN^\prime} N_{J}^{-1}\int dg ~{\tilde{\tilde W}}_1(g; JNN^\prime)
{\tilde{\tilde W}}_2(g; JN^\prime N)={\rm Tr}({\hat \rho}_1{\hat \rho}_2). 
\label{81}
\end{equation}

The conclusion we can draw, in interesting contrast to the Cartesian and
Abelian cases, is this. In order to be able to recover the marginal
probability distributions $<g|{\hat \rho}|g>,~<JMN|{\hat \rho}|JMN>$ in
natural ways and also to have simple transformation behaviours under both left
and right translations on $G$, we need to define the Wigner distribution as in
eq.$(\ref{65},\ref{77})$ with independent arguments 
$g~J~M~N~M^\prime~N^\prime$. However, this object captures information
contained in ${\hat \rho}$ in an over complete manner, since ${\hat \rho}$ is
in fact completely determined in principle already by 
$ {\tilde W}(g; JMM^\prime )$ ( or ${\tilde{\tilde W}}(g; JNN^\prime)$). All
this is traceable to the fact that for non Abelian $G$, the UIR's are in
general multidimensional, so the concept of `momentum eigenstate' is also a
multidimensional set of vectors.

\section {Recovery of the Cartesian and angle-angular momentum cases, and the 
$SU(2)$ case}

We now indicate briefly how the known earlier results of Sections II, III can
be immediately recovered from the definitions of the previous Section. The
expression $(\ref{65})$ for the Wigner distribution 
$W(g;~JMN~M^\prime N^\prime)$ uses the function $s(g^\prime,g^{\prime\prime})$
depending symmetrically on the group elements $g^\prime,g^{\prime\prime}$, and
is itself a group element obeying the conditions in $(\ref{66})$. For the 
case of a compact simple Lie group $G$ with nontrivial Cartan-Killing metric 
and associated geodesics, we have found a solution for 
$s(g^\prime,g^{\prime\prime})$ in terms of the mid point rule. If however 
$G$ is Abelian we can directly give an elementary solution for 
$s(g^\prime,g^{\prime\prime})$ not using the geodesic construction at all. 

For Cartesian quantum mechanics we have $G={\cal R}^n$, which is
Abelian. Consequently each UIR of $G$ is one-dimensional and corresponds to a
definite numerical momentum vector:
\begin{equation}
{\underline q} \in G \longrightarrow e^{i{\underline q}\cdot {\underline p}/\hbar}, 
{\underline p} \in {\cal R}^n. 
\end{equation}

We can regard the continuous vector 
${\underline p} \in {\cal R}^n$ (actually dual 
to $G$, the space of characters) as the analogue of the label $J$ of the
previous section, and as each UIR is one dimensional there is no need and no
room for the labels $M~N~M^\prime~N^\prime$. If we present the usual 
definition $(\ref{11})$ in the form 
\begin{eqnarray}
W({\underline q}, {\underline p}) &=& (2\pi\hbar)^{-n}\int_{{\cal R}^n} 
d^n q^\prime \int_{{\cal R}^n} d^n q^{\prime\prime}
~~\delta^{(n)}({\underline q}-{\underline s}({\underline q^\prime}, {\underline
  q^{\prime\prime}})) <{\underline q^{\prime\prime}}|{\hat \rho}|
{\underline q^\prime}> \nonumber \\
&&~~~~~~~~~~~~~~~~~\exp(i{\underline q^\prime}\cdot {\underline p}/\hbar) 
~~\exp(-i{\underline q^{\prime\prime}}\cdot {\underline p}/\hbar)\nonumber \\
{\underline s}({\underline q^\prime}, {\underline
  q^{\prime\prime}})&=&\frac{1}{2}({\underline q^\prime}+
{\underline q^{\prime\prime}}),
\end{eqnarray}
we see that all the conditions $(\ref{66})$ are indeed obeyed and this
familiar case is seen to be a special case of our general construction. 

The key point is that our construction of the Wigner distribution only depends
on finding the group element $s( g^\prime, 
 g^{\prime\prime})$. We may use the geodesic construction if it is
available, but can use any other method if a metric on $G$ and geodesics are
not available. 

Turning to the compact case $G=SO(2)$, this is again Abelian, so each UIR is
one dimensional: 
\begin{equation}
\theta \in G \longrightarrow e^{im\theta},~~m=0, \pm 1, \pm 2,\cdots  . 
\end{equation}
We can now write the Wigner distribution $(\ref{26})$ as 
\begin{eqnarray}
W(\theta,m) &=& \frac{1}{2\pi}\int_{-\pi}^{\pi} 
d\theta^\prime \int_{-\pi}^{\pi} d\theta^{\prime\prime}
~~\delta(\theta- s(\theta^\prime, \theta^{\prime\prime})) 
<\theta^{\prime\prime}|{\hat \rho}| \theta^\prime> \nonumber \\
&&~~~~~~~~~~~~~~~~~\exp(im\theta^\prime) 
~~\exp(-im\theta^{\prime\prime})\;,\nonumber \\
s(\theta^\prime, \theta^{\prime\prime})&=&\frac{1}{2}(\theta^\prime+
\theta^{\prime\prime})~ {\rm mod}~2\pi,
\end{eqnarray}
and again see that it falls into our general pattern. 

 Finally in this Section we present briefly the structure and some
significant features of Wigner functions in the case 
$G=SU(2)$, in a sense the simplest yet archetypal compact non
Abelian Lie group.  Here the method of geodesics is essential 
for the construction.  We recall very rapidly the basic definitions 
and notations concerning $SU(2)$, emphasizing the four
dimensional geometric aspects available in this case

The defining representation of $SU(2)$ is via $2\times 2$ unitary 
unimodular matrices, which leads immediately to the 
identification of the group manifold with $S^3$, the  real
unit sphere in four dimensional Euclidean space ${\cal R}^4$.
We shall exploit this way of picturing $SU(2)$.  We denote
group elements in the abstract by $a,b,a^{\prime},b^{\prime},
\ldots,$ these symbols also standing for points on $S^3$:

        \setcounter{equation}{4}
         \begin{eqnarray}
     a&=& (a_{\mu})\in S^3,\;\mu=0, 1,2,3;\nonumber\\
     a_{\mu} a_{\mu}&=& a^2_0 + \underline{a}\cdot
     \underline{a}= 1
     \end{eqnarray}  

\noindent
The `spatial' part $(a_1,a_2,a_3)$ of $(a_{\mu})$ is denoted 
by $\underline{a}$.  Inverses and products of group elements 
are denoted by $a^{-1}, ab$ respectively.  (The group
element $ab$ is to be carefully distinguished from the four
vector inner product $a\cdot b$ which is a real number).
Then in the defining representation the matrix corresponding
to $a\in SU(2)$ is
     \begin{eqnarray}
     u(a) &=& a_0\cdot I -i\; \underline{a}\cdot
     \underline{\sigma}
     = \pmatrix{\lambda&\mu\cr-\mu^*&\lambda^*},\nonumber\\
      \lambda &=& a_0 -i a_3,\; \mu=-(a_2 +ia_1).
      \end{eqnarray}  

\noindent
Here $\underline{\sigma}=(\sigma_1, \sigma_2, \sigma_3)$ are 
the Pauli matrices.  The inverse arises by reversing the 
sign of $\underline{a}$:
     \begin{eqnarray}
     u(a)^{-1} = u(a^{-1}) = u(a_0,-\underline{a}) =
     a_0\cdot I + i\underline{a}\cdot \underline{\sigma}.
     \end{eqnarray}

\noindent
The group multiplication law is subsumed in the
description of left and right translations, each of which
is realised by elements of $SO(4)$:
     \begin{mathletters}
     \begin{eqnarray}
     u(a)u(b)&=& u(ab) = u(L(a)b),\nonumber\\
     L(a)&=&\pmatrix{
     a_0&-a_1&-a_2&-a_3\cr
     a_1&a_0&-a_3&a_2\cr
     a_2&a_3&a_0&-a_1\cr
     a_3&-a_2&a_1&a_0}
     \in SO(4);\\
     &&\nonumber\\
     u(b)u(a^{-1})&=& u(ba^{-1}) = u(R(a)b),\nonumber\\
     R(a)&=&\pmatrix{
     a_0&a_1&a_2&a_3\cr
     -a_1&a_0&-a_3&a_2\cr
     -a_2&a_3&a_0&-a_1\cr
     -a_3&-a_2&a_1&a_0}
      \in SO(4);\\
     &&\nonumber\\
     L(a) L(b)&=& L(ab),\;R(a) R(b) = R(ab);\nonumber\\
     L(a) R(b) &=& R(b) L(a) .    
     \end{eqnarray}
     \end{mathletters}
   
\noindent
Each of these mutually commuting sets $\{L(a)\}, \{R(a)\}$ 
faithfully represents $SU(2)$ via $SO(4)$ matrices.  The
only common elements correspond to $a=(\pm 1,\underline{0})$
leading to the two matrices $Z_2=\{\pm I\}$ within
$SO(4)$.  This leads to the familiar statement
     \begin{eqnarray}
     SO(4) = SU(2)\times SU(2)/Z_2 .
\label{82}     
\end{eqnarray}

\noindent
We mention these details since the general covariance
requirements in eqns.$(\ref{63c},\ref{63d})$ require them.

The relation to the Euler angles parametrisation is
given by
     \begin{eqnarray*}
     u(a)&=&e^{-\frac{i}{2}\alpha\sigma_3}
     e^{-\frac{i}{2}\beta\sigma_2}
     e^{-\frac{i}{2}\gamma\sigma_3},\nonumber\\
     a_0-ia_3&=& \cos \beta/2\;e^{-i(\alpha+\gamma)/2},
     \nonumber\\
     a_2+ia_1&=& \sin \beta/2\;e^{i(\gamma-\alpha)/2},
     \end{eqnarray*}

    \begin{eqnarray}
    0\leq \alpha \leq 2\pi,\;0\leq\beta\leq \pi,\;
    0\leq\gamma\leq 4\pi .
    \end{eqnarray} 

\noindent
We can regard $\alpha,\beta,\gamma$ as angular coordinates over
$S^3$, though because of the occurrence of half angles they
are not quite the natural generalisation of spherical
polar angles from $S^2$ to $S^3$.  The invariant line 
element $(ds)^2$ on $S^3$, the invariant normalised volume
element $da$ on $SU(2)$, and the element of solid angle
$d\Omega(a)$ on $S^3$ can all be easily worked out:
        \begin{mathletters}
        \begin{eqnarray}
        (ds)^2 &=& da_{\mu} da_{\mu} = |d(a_0-ia_3)|^2
        +|d(a_2+ia_1)|^2\nonumber\\
        &=&\frac{1}{4}((d\alpha)^2 + (d\beta)^2 +
        (d\gamma)^2 +2\cos \beta\; d\alpha \;d\gamma);\\
        da&=& \frac{1}{2\pi^2} d\Omega(a) =
        \frac{1}{16\pi^2} d\alpha\; \sin \beta\; d\beta 
        \;d\gamma.
        \end{eqnarray}
        \end{mathletters}

It is clear that the above line element on $S^3$ is the one 
induced from the Euclidean line element in ${\cal R}^4$, hence 
the corresponding  geodesics are `great circle arcs'.
Such arcs are carried into one another by both left and 
right $SU(2)$ translations - the $SO(4)$ invariance of 
$(ds)^2$ along with eqn.($\ref{82})$ make this obvious.  Therefore,
if $a,b$ are any two points of $S^3$ (any two elements of 
$SU(2)$) which are not diagonally opposite one another
$(u(a)\neq -u(b))$, the (shorter) geodesic connecting
them is the affinely parametrised curve
  
     \[a(\theta) =a \cos \theta + (b-a\;a\cdot b) \sin 
     \theta/\sqrt{1-(a\cdot b)^2},\]
     \begin{eqnarray}
     0\leq \theta \leq \theta_0 = \cos^{-1} (a\cdot b)
     \in [0,\pi).
     \end{eqnarray}

\noindent
Along this geodesic we have $(ds)^2=(d\theta)^2$, and
the mid point is given by
     \begin{eqnarray}
     a\left(\frac{\theta_0}{2}\right) =
     (a+b) /\sqrt{2(1+a\cdot b)}.
\label{83}     
\end{eqnarray}
 
\noindent
which is geometrically obvious.

The Dirac delta function accompanying the volume element 
$da$ on $SU(2)$ may be written as $\delta(a,b)$ involving
two group elements, or in a more compact form as
$\delta(a^{-1}b)$.  Its properties are summarised by
     \begin{eqnarray*}
     \int\limits_{SU(2)} db\;\delta(a,b)\;f(b)
     \equiv \int\limits_{S^3}\frac{d\Omega(b)}{2\pi^2}
     \delta(a,b)\;f(b) = f(a) ,
     \end{eqnarray*}

     \begin{eqnarray}
     \mbox{ie}\;\; \int\limits_{S^3}\;d\Omega(b)\;
     \delta(a,b)\;f(b) = 2\pi^2 f(a) .
     \end{eqnarray}

\noindent
for suitable test functions $f(b)$. We can equally well regard
$\delta(a,b)$ as a delta function on $SU(2)$ or on 
$S^3$.  In terms of Euler angles we have:  
     \begin{eqnarray}
     a\rightarrow (\alpha,\beta, \gamma)&,&
     b\rightarrow (\alpha^{\prime},\beta^{\prime},
     \gamma^{\prime}):\nonumber\\
     \delta(a,b) &=& 16\pi^2 \;\delta(\alpha^{\prime}-
     \alpha) \;\delta(\beta^{\prime}-\beta)\;
     \delta(\gamma^{\prime}-\gamma)/\sin\beta .
     \end{eqnarray}

The last item in this resume concerns the matrices 
$D^j_{mm^{\prime}}(a)$ representing $SU(2)$ elements in the 
various UIR's.  The ranges of the UIR label $j$ and
magnetic quantum numbers $m,m^{\prime}$ are, as usual,
$j=0,1/2,1,3/2,\ldots,\;m,m^{\prime}=j,j-1,\ldots,-j$.
Then with canonical basis vectors $|j m>$ in the $j^{\mbox{th}}$
UIR and hermitian generators $J_1,J_2,J_3$ we have from
the quantum theory of angular momentum \cite{gamma}:
     \begin{eqnarray}
     D^j_{mm^{\prime}}(a) &=& \langle jm|e^{-i\alpha J_3} 
     e^{-i\beta J_2} e^{-i\gamma J_3}|jm^{\prime}\rangle
     \nonumber\\
     &=&e^{-i m \alpha - i m^{\prime} \gamma} 
     d^j_{mm^{\prime}}(\beta),\nonumber\\
     d^j_{mm^{\prime}}(\beta) &=&
     \langle jm|e^{-i\beta J_2}|jm^{\prime}\rangle\nonumber\\
     &=&\sqrt{\frac{(j+m^{\prime})!(j-m^{\prime})!}
     {(j+m)! (j-m)!}} \left(\sin \frac{\beta}{2}\right)
     ^{m^{\prime}-m}  \left(\cos\frac{\beta}{2}\right)
     ^{m^{\prime}+m} P_{j-m^{\prime}}^{(m^{\prime}-m,
     m^{\prime}+m)} (\cos\beta),
     \end{eqnarray}

\noindent
where the $P$'s are the Jacobi polynomials.  The 
orthogonality and completeness properties of these
$D$-functions are:
     \begin{eqnarray*}
     \int\limits_{SU(2)} da\;D^j_{mm^{\prime}}(a)^*
     D^{j^{\prime}}_{m^{\prime\prime}m^{\prime\prime\prime}}
     (a) = (2j+1)^{-1}\; \delta_{jj^{\prime}}\;
      \delta_{mm^{\prime\prime}}\;\delta_{m^{\prime}
      m^{\prime\prime\prime}};
     \end{eqnarray*}
     \begin{eqnarray}
     \sum_{j=0,1/2,1}\;\sum^j_{m,m^{\prime}=-j}
     (2j+1)D^j_{mm^{\prime}}(a)\;
     D^j_{mm^{\prime}}(b)^* = \delta(a,b)=\delta(a^{-1}b).
     \end{eqnarray}

With these details in place we can proceed to the definition 
of the Wigner function.  The Hilbert space of Schr\"odinger
wave functions is 
     \begin{eqnarray}
     {\cal H} = L^2(SU(2))&=&\{\psi(a) \in C|a\in SU(2),
     \nonumber\\
     &&\left.\parallel\psi\parallel^2 = \int\limits
     _{SU(2)}\;da\;|\psi(a)|^2 < \infty\right\}.
     \end{eqnarray}
\noindent
Given $\psi\in{\cal H}$, the corresponding Wigner function is 
obtained by specialising eqns.$(\ref{65},\ref{67},\ref{75})$ to this case 
and using eqn.$(\ref{83})$ above:
     \begin{eqnarray}
     W(a;jmn\;m^{\prime}n^{\prime})&=&
     \frac{(2j+1)}{4\pi^4} \int\limits_{S^{3}}\;d\Omega
     (a^{\prime}) \int\limits_{S^{3}} \;d\Omega 
     (a^{\prime\prime})\;\delta\left(a,\frac{a^{\prime}+
     a^{\prime\prime}}{\sqrt{2(1+a^{\prime}\cdot a^{\prime
     \prime})}}\right)\times\nonumber\\
     &&D^j_{mn}(a^{\prime}) \psi(a^{\prime})^*\;
     D^j_{m^{\prime}n^{\prime}}(a^{\prime\prime})^*\;
     \psi(a^{\prime\prime}).
\label{84}    
\end{eqnarray}

\noindent
The occurrence of the mid point of the geodesic from $a^{\prime}$
to $a^{\prime\prime}$ within the delta function is to be noted.  
We see immediately that the marginals are properly reproduced:
     \begin{mathletters}
\label{85}
     \begin{eqnarray}
     \int\;da\;W(a;jmn\;m^{\prime}n^{\prime})&=&
     \psi_{jm^{\prime}n^{\prime}}\psi^*_{jmn},\nonumber\\
     \psi_{jmn}&=&\frac{\sqrt{2j+1}}{2\pi^2}\int\;d\Omega(a)
      \;D^j_{mn}(a)^*\;\psi(a);\\
     \sum_{jmn}&&W(a;jmn\;mn) = |\psi(a)|^2 .
     \end{eqnarray}
     \end{mathletters} 

Since the integrations involved in eqn.$(\ref{84})$ are nontrivial, 
we limit ourselves to pointing out some qualitative features 
of the $SU(2)$ Wigner function $(\ref{84})$ which distinguish it 
from the Cartesian case as well as from earlier treatments
of this problem:

(a)The appearance of all the arguments  $ a\; jmn\;m^{\prime}
n^{\prime}$ in the Wigner function is essential to be able
to satisfy the covariance laws $(\ref{63c},\ref{63d})$ under independent 
left and right $SU(2)$ translations, and to reproduce the
`configuration space' and `momentum space' marginal
probability distributions as in eqn.$(\ref{85})$.  In this
respect the situation is markedly different from
earlier approaches to the $SU(2)$ Wigner function problem \cite{3},
where attention was limited to states within some fixed 
(finite dimensional) UIR of $SU(2)$ and the density matrix 
was expanded in the complete set of unit tensor operators 
within that UIR.

(b)  If we consider as an idealised limit the case of
$\psi(a)$ becoming a `position eigenstate', the 
Wigner function simplifies as follows:
     \begin{eqnarray}
     \psi(a)\rightarrow \delta(a,a^{(0)}):&&\nonumber\\
     W(a; jmn\;m^{\prime}n^{\prime})&=&
     \frac{(2j+1)}{4\pi^4}\;\delta(a,a^{(0)})
    \;D^j_{mn}(a^{(0)})\;D^j_{m^{\prime}n^{\prime}}
    (a^{(0)})^* .
    \end{eqnarray}

\noindent
This retains a dependence on the `momentum variables' 
$jmn\;m^{\prime}n^{\prime}$.  This is in contrast to 
the (one-dimensional) Cartesian case where from
eqn.$(\ref{10})$ we find:
     \begin{eqnarray}
      \psi(q)\rightarrow \delta(q-q_0)\;:\;
     W(q,p) =\frac{1}{h}\;\delta(q-q_0),
\label{86}     
\end{eqnarray}

\noindent
showing no $p$ dependence.

(c)  Similarly if we consider $\psi(a)$ to be a (normalised)
linear combination of $D^{j_{0}}_{m_{0}n_{0}}(a)$ over 
$m_{0}\;n_{0}$ for some fixed $j_0$, the Wigner function
has a nontrivial dependence on all its arguments, and
\underline{in particular is generally nonvanishing for 
$j\neq j_0$}.  In the Cartesian case, in contrast, we 
have, similar to eqn.$(\ref{86})$:
     \begin{eqnarray}
     \psi(q)\rightarrow\frac{1}{h^{1/2}}\;e^{ip_{0}q}\;:\;
    W(q,p)=\frac{1}{h}\;\delta(p-p_0),
    \end{eqnarray}

\noindent
concentrated at $p=p_0$ and independent of $q$.

All these features can be attributed to the non Abelian
nature of $SU(2)$.

\section{Concluding remarks}

We have discussed the problem of setting up Wigner distributions for the
states of a quantum system whose configuration space is a general non Abelian 
Lie group $G$, and have given a complete solution for the case that $G$ is
compact. Many new features compared to the familiar Abelian case where 
$G={\cal R}^n$ have appeared. For emphasis we repeat some of them here: while
the classical phase space $T^*G$ associated to $G$ already brings in
interesting structural aspects, in the quantum case the Wigner distribution is
not a function defined on the classical $T^*G$. Instead it is a function of a 
`classical unquantised' group element $g\in G$ playing the role of coordinate
variable, and `quantised momenta' consisting of labels 
$J~M~N~M^\prime~N^\prime$ associated with all the UIR's of $G$. The analogues
of the familiar Heisenberg canonical commutation relations are now much more
intricate, and the ideas of `momentum eigenstates'  and `momentum eigenvalues'
have to be understood with some care. While the function 
$W(g;~JMN~M^\prime N^\prime)$ associated with a given ${\hat \rho}$ 
transforms nicely under left and right group actions, and reproduces the 
marginal probability distributions satisfactorily, it describes 
${\hat \rho}$ in an over complete manner. 

The points of view of the present work suggest that we also consider quantum
systems whose covariance group is a given Lie group $G$, even if $G$ is not
the configuration space. These arise naturally if the configuration space is a
coset space $G/H$, where $H$ is some Lie subgroup of $G$. In that case there
is only one (say left) action of $G$ on $G/H$, rather than two independent 
mutually commuting actions. Action by $G$ remains significant, and we would
like to set up Wigner distributions for wave functions belonging to 
$L^2(G/H)$. Such UR's of $G$ are typically much smaller than the regular
representation.   

Going beyond coset space representations, we have yet other
physically interesting cases typified for example by the
Schwinger oscillator representation of $SU(2)$.  Similar
constructions are easily made for $SU(3)$ etc.\cite{delta}. These are not
representations on spaces $L^2(G/H)$ for any choice of $H$; yet
because of their use in various physical problems it is 
worthwhile to be able to set up Wigner distributions for them 
too.  

We intend to examine some of these problems elsewhere. 

\newpage
\def\theequation{A.\arabic{equation}}
\appendix{\bf{Appendix A}: The Regular Representation and associated 
structures}
\setcounter{equation}{0}

We assemble here some familiar facts concerning the regular
representation of a compact Lie group, to settle notations and
as preparation for setting up further operator structures.  We
know that the Lie group $G$ under consideration possesses a left
and right translation and inversion invariant volume element,
$dg$ say, so that the integral of a (complex valued) function
$f(g)$ over $G$ has the properties \begin{eqnarray}
\int\limits_{G}\; dg \;f(g) = \int\limits_G\; dg
\;\left(f(g^{-1}) \;\mbox{or}\; f(g^{\prime} g)\;\mbox{or}\;
f(gg^{\prime})\right) , \end{eqnarray}

\noindent
where $g^{\prime}$ is any fixed element in $G$.  For the compact
case we normalise $dg$ so that \begin{eqnarray} \int\limits_G\;
dg = 1 \end{eqnarray}

\noindent
With such a measure the carrier space for the unitary regular
representation of $G$ is the Hilbert space ${\cal H}=L^2(G)$
defined as in eqn.$(\ref{55})$: \begin{eqnarray} {\cal H} =
\{\psi(g)\; \in \; {\cal C}|\parallel\psi\parallel^2 =
\int\limits_G \;dg |\psi(g)|^2 < \infty\} \end{eqnarray}

\noindent
A Dirac delta function can be defined with suitable invariance
properties: \begin{eqnarray} \int\limits_G\; dg\; f(g)\;
\delta(g) = \int\limits_G\; dg\; f(g)\;
\delta\left(g^{-1}\right) &=& f(e) ,\nonumber\\ \int\limits_G
\;dg\; f(g)\; \delta\left(g g^{\prime -1}\;\mbox{or}\; g^{-1}
g^{\prime}\;\mbox{or}\;g^{\prime -1}g\;\mbox {or}\;g^{\prime}
g^{-1}\right) &=&       f(g^{\prime}) .  \end{eqnarray}

\noindent
We can introduce a convenient set of ideal basis vectors for
${\cal H}$ such that the wave function $\psi(g)$ is the overlap
of $|\psi\rangle$ with one of these: 
\begin{eqnarray} \psi(g)&=&
\langle g|\psi\rangle ,\nonumber\\ \langle g^{\prime}|g\rangle
&=& \delta(g^{\prime} g^{-1}) ,\nonumber\\ \int\limits_G dg
|g\rangle\langle g|&=& 1\;\mbox{on}\;{\cal H} .  
\label{A.5}
\end{eqnarray}
   
The group $G$ can be unitarily represented on ${\cal H}$ in two
mutually commuting ways, by left or by right translations.  We
denote the corresponding operators by $V(g), \tilde{V}(g)$ and
define them by: 
\begin{eqnarray} 
V(g^{\prime})|g\rangle &=&
|g^{\prime} g\rangle , \nonumber\\
\tilde{V}(g^{\prime})|g\rangle &=& |g g^{\prime -1}\rangle .
\label{A.6}
\end{eqnarray}

\noindent
Both of them are unitary and obey the composition and
commutation relations \begin{eqnarray} V(g_2) V(g_1) &=& V(g_2
g_1), \nonumber\\ \tilde{V}(g_2) \tilde{V}(g_1)&=& \tilde{V}(g_2
g_1),\nonumber\\ V(g_1) \tilde{V}(g_2) &=& \tilde{V}(g_2) V(g_1)
.  \end{eqnarray}

\noindent
On wave functions the effects are as given in eqn.$(\ref{56})$:
\begin{eqnarray} (V(g^{\prime})\psi)(g) &=& \psi(g^{\prime -1}g)
,\nonumber\\ (\tilde{V}(g^{\prime}) \psi) (g) &=& \psi(g
g^{\prime}) .  \end{eqnarray}

These are infinite dimensional reducible UR's of $G$; and in the
compact case, according to the Peter-Weyl theorem, each of them
contains every UIR of $G$ as often as its dimension.  Motivated
by the notations in the case of $SU(2)$, we shall use symbols
$J,J^{\prime},J_1,J_2,\ldots$ to label the various UIR's of $G$
(some of which may not be faithful); so in fact $J$ stands for
several independent discrete or quantised labels, as many as the
rank of $G$.  Within the $J$th UIR, in some chosen orthonormal
basis, we label  rows and columns by indices $M,N,M^{\prime}
N^{\prime},\ldots$.  Once again each of these stands for a
collection of discrete labels: for instance the eigenvalues of
as many commuting generators as the rank of $G$, plus further
eigenvalues of chosen commuting nonlinear polynomials in the
generators.  In the $J$th UIR, we write ${\cal D}^J_{MN}(g)$ for the
unitary representation matrices.  These obey composition,
orthogonality and completeness relations: 
\begin{eqnarray} 
\sum\limits_{M^{\prime}} {\cal D}^J_{MM^{\prime}}
(g^{\prime}) {\cal D}^J_{M^{\prime}N} (g) &=& 
{\cal D}^J_{MN}(g^{\prime}g) ;\label{A.9a}\\
\int\limits_G dg\;{\cal D}^{J^{\prime}}_{M^{\prime}N^{\prime}} 
(g)^*{\cal D}^J_{MN}(g) &=& \delta_{J^{\prime}J}
\delta_{M^{\prime}M}\delta_{N^{\prime}N} /N_J ;\label{A.9b}\\
\sum\limits_{JMN} N_J {\cal D}^J_{MN} (g) 
{\cal D}^J_{MN}(g^{\prime})^* &=&
\delta\left(g^{-1} g^{\prime}\right) .  
\label{A.9c}
\end{eqnarray}

\noindent
Here $N_J$ is the dimension of the $J$th UIR.  With the help of
these matrices we can introduce another orthonormal basis for
${\cal H}$ which explicitly accomplishes the simultaneous
reduction of both UR's $V(\cdot), \tilde{V}(\cdot)$ into
irreducibles.  These basis vectors and their main properties
are: 
\begin{eqnarray} 
|JMN\rangle &=& N^{1/2}_J \int\limits_G
dg {\cal D}^J_{MN}(g)|g\rangle , \nonumber\\ 
\langle g| JMN\rangle &=&
N^{1/2}_J {\cal D}^J_{MN} (g) ;\nonumber\\ \langle J^{\prime}
M^{\prime}N^{\prime}| JMN\rangle &=& \delta_{J^{\prime}J}
\delta_{M^{\prime}M} \delta_{N^{\prime}N} ; \nonumber\\
\sum\limits_{JMN} |JMN \rangle\langle JMN|&=& 1\;
\mbox{on}\;{\cal H} .  
\label{A.10}
\end{eqnarray}

\noindent
Under action by $V(\cdot), \tilde{V}(\cdot)$ they transform
among themselves conserving $J$: 
\begin{eqnarray} V(g)
|JMN\rangle &=& \sum\limits_{M^{\prime}} {\cal D}^J_{MM^{\prime}}
\left(g^{-1}\right) |JM^{\prime}N\rangle ,\nonumber\\
\tilde{V}(g)|JMN\rangle &=& \sum\limits_{N^{\prime}}
{\cal D}^J_{N^{\prime}N}(g) |JMN^{\prime}\rangle .  
\label{A.11}
\end{eqnarray}

\noindent
Therefore in $|JMN\rangle$ the index $N$ counts the multiplicity
of occurrence of the $J$th UIR in the reduction of $V(\cdot)$,
and the index $M$ performs a similar function in the reduction
of $\tilde{V}(\cdot)$.  A general $|\psi\rangle$ can be expanded
in either basis and we have: 
\begin{eqnarray} |\psi\rangle =
\int\limits_G\; dg\; \psi(g) |g\rangle &=& \sum\limits_{JMN}
\psi_{JMN} |JMN\rangle ,\nonumber\\ \psi_{JMN} = \langle
JMN|\psi\rangle &=& N_J^{1/2} \int\limits_G\; dg \;{\cal D}^J_{MN}
(g)^* \psi(g) ,\nonumber\\ \parallel\psi\parallel^2 =
\int\limits_G\; dg\; |\psi(g)|^2 &=& \sum\limits_{JMN}
|\psi_{JMN}|^2 .  
\label{A.12}
\end{eqnarray}

Towards getting projections onto individual vectors
$|JMN\rangle$ we set up the `Fourier components' of $V(\cdot)$
and $\tilde{V}(\cdot)$ as follows: 
\begin{eqnarray} P_{JMN}&=&
N_J \int\limits_G\;dg\;{\cal D}^J_{MN} (g) V(g) ,\nonumber\\
\tilde{P}_{JMN}&=& N_J \int\limits_G\;dg\;{\cal D}^J_{MN} (g^{-1})
\tilde{V}(g) .  \end{eqnarray}

\noindent
With respect to hermitian conjugation the indices get
interchanged: \begin{eqnarray} P^{\dag}_{JMN} =
P_{JNM},\;\tilde{P}_{JMN}^{\dag} = \tilde{P}_{JNM} ;
\end{eqnarray}

\noindent
and their composition and multiplication laws are:
\begin{eqnarray} P_{J^{\prime}M^{\prime}N^{\prime}} P_{JMN} &=&
\delta_{J^{\prime}J} \delta_{N^{\prime}M} P_{JM^{\prime}N}
,\nonumber\\ \tilde{P}_{J^{\prime}M^{\prime}N^{\prime}}
\tilde{P}_{JMN} &=& \delta_{J^{\prime}J} \delta_{M^{\prime}N}
\tilde{P}_{JMN^{\prime}} , \nonumber\\ P_{JMN}
\tilde{P}_{J^{\prime}M^{\prime}N^{\prime}} &=&
\tilde{P}_{J^{\prime}M^{\prime}N^{\prime}} P_{JMN} .
\end{eqnarray}

\noindent
Their actions on the two complementary bases for ${\cal H}$ are
immediate: \begin{eqnarray} P_{J^{\prime}M^{\prime}N^{\prime}}
|JMN\rangle &=& \delta_{J^{\prime}J} \delta_{N^{\prime}M}
|JM^{\prime}N\rangle      ,\nonumber\\
\tilde{P}_{J^{\prime}M^{\prime}N^{\prime}} |JMN\rangle &=&
\delta_{J^{\prime}J} \delta_{M^{\prime}N} |JMN^{\prime}\rangle ;
\nonumber\\ P_{JMN} |g\rangle &=& N^{1/2}_J
\sum\limits_{N^{\prime}}
{\cal D}^J_{N^{\prime}N}(g^{-1}) |JMN^{\prime}\rangle , \nonumber\\
\tilde{P}_{JMN}|g\rangle &=& N^{1/2}_J \sum\limits_{M^{\prime}}
{\cal D}^J_{MM^{\prime}}(g^{-1})|JM^{\prime}N\rangle .  
\end{eqnarray}

\noindent
Therefore the projections onto $|JMN\rangle$ are
\begin{eqnarray} |JMN\rangle\langle JMN| = P_{JMM}
\tilde{P}_{JNN}, \end{eqnarray}

\noindent
and we have the completeness identities \begin{eqnarray}
\sum\limits_M \;P_{JMM} &=& \sum\limits_M\;\tilde{P}_{JMM}
,\nonumber\\ \sum\limits_{JM}\;P_{JMM} &=& \sum\limits_{JM}\;
\tilde{P}_{JMM} = 1\;\mbox{on}\;{\cal H}.  \end{eqnarray}

Now we proceed to a construction of certain operators directly
relevant to the Wigner distribution problem.  Here we will be
guided by analogy to what is done for the (one degree of
freedom) $\hat{q}-\hat{p}$ pair and the
$\widehat{\theta}-\widehat{M}$ pair, as recounted in Sections II 
and III. In these cases we know that the unitary Weyl exponentials
$U(\sigma)=\exp(i\sigma\hat{q}), V(\tau)=\exp(-i\tau\hat{p})$
and $U(n)=\exp(in\widehat{\theta}),
V(\tau)=\exp(-i\tau\widehat{M})$ play important roles.  It is
seen that it is natural here to regard $\sigma(n)$ as a typical
eigenvalue of $\hat{p}(\widehat{M})$ and $\tau$ (in ${\cal R}$ or
in $(-\pi,\pi)$) as a typical eigenvalue of
$\hat{q}(\widehat{\theta})$.  Now the operator $V(\tau)$ has
been generalised in the Lie group situation to the
\underline{two} families $V(g), \tilde{V}(g)$.  These are indeed
exponentials of the ``momentum operators'':  if the group
element $g$ is expressed as the exponential of an element in
$\underline{G}$, then $V(g)$ and $\tilde{V}(g)$ are
corresponding exponentials in their generators $(\ref{57})$ obeying
$(\ref{53})$: 
\begin{eqnarray} 
g = \exp\left(\tau^r\;e_r\right) :
V(g)&=&\exp\left(-
i\;\tau^r\widehat{J}_r\right) ,\nonumber\\ \tilde{V}(g) &=&
\exp\left(-i\;\tau^r \widehat{\tilde{J}}_r\right).
\end{eqnarray}

\noindent
With $\tau^r$ as coordinates for $g$,  these are precisely
exponentials in momenta.  To generalise $U(\sigma), U(n)$ we
recall on the other hand that now a typical ``momentum
eigenvalue'' is the collection of quantum numbers $JMN$
associated with a subspace of ${\cal H}$ supporting a UIR of
the $\widehat{J}_r, \widehat{\tilde{J}}_r$.  This suggests that
the generalisation of $U(\sigma), U(n)$ must be an operator
diagonal in the ``coordinate'' or $|g\rangle$ basis, and
labelled by $JMN$: it must be a `function of the coordinates'
alone.  Based on this reasoning, we define operators $U_{JMN}$
by: 
\begin{eqnarray} 
U_{JMN} |g\rangle &=& {\cal D}^J_{MN}(g) |g\rangle
\nonumber\\ \langle g| U_{JMN} &=& {\cal D}^J_{MN}(g) \langle g| .
\label{A.20}
\end{eqnarray}

\noindent
Their adjoints are also diagonal in this basis: 
\begin{eqnarray}
U_{JMN}^{\dag} |g\rangle &=& {\cal D}^J_{MN}(g)^* |g\rangle
,\nonumber\\ \langle g| U_{JMN}^{\dag} &=& 
{\cal D}^J_{MN}(g)^* \langle
g| , \end{eqnarray}

\noindent
and unitarity is expressed in a matrix sense: 
\begin{eqnarray}
\sum\limits_M U_{JMN}^{\dag} U_{JMN^{\prime}} = \sum\limits_M
U_{JNM}^{\dag} U_{JN^{\prime}M} = \delta_{N^{\prime}N}\cdot 1
\;\mbox{on}\;{\cal H} \end{eqnarray}

\noindent
Being simultaneously diagonal, the commutators vanish:
\begin{eqnarray} \protect[U_{JMN},
U_{J^{\prime}M^{\prime}N^{\prime}}\protect] = \protect[U_{JMN},
U_{J^{\prime}M^{\prime}N^{\prime}}^{\dag} \protect] = 0.
\end{eqnarray}
  
\noindent
Completeness of the ${\cal D}$-functions ${\cal D}^J_{MN}(g)$ as 
expressed in eqn.$(\ref{A.9c})$ now means that the operators 
$\{U_{JMN}\}$ form a
(linear) basis for the commutative algebra $\widehat{{\cal A}}$.
In fact the map $(\ref{51})$ from the classical algebra ${\cal A}$ to
the quantised $\widehat{{\cal A}}$ can be made explicit:
\begin{eqnarray} f\;\in \;{\cal A} : f(g) &=&
\sum\limits_{JMN} f_{JMN}      {\cal D}^J_{MN}(g)\longrightarrow
\nonumber\\ \hat{f}&=& \sum\limits_{JMN} f_{JMN}
U_{JMN}\;\in\;\widehat {{\cal A}} .  \end{eqnarray}

\noindent
The relations connecting $\{U_{JMN}\}$ to $V(\cdot),
\tilde{V}(\cdot)$ are easily worked out: \begin{eqnarray} V(g)
U_{JMN} V(g)^{-1} &=& \sum\limits_{M^{\prime}} 
{\cal D}^J_{MM^{\prime}}
(g^{-1}) U_{JM^{\prime}N}, \nonumber\\ \tilde{V}(g) U_{JMN}
\tilde{V}(g)^{-1} &=& \sum\limits_{N^{\prime}} 
{\cal D}^J_{N^{\prime}N}
(g) U_{JMN^{\prime}}.  \end{eqnarray}

\noindent
What remains are expressions for the product of two $U$'s, and
the action of a $U$ on $|JMN\rangle$.  For both these, the
Clebsch-Gordan coefficients for $G$ have to be brought in.

Let the reduction of the direct product of the $J$th and
$J^{\prime}$th UIR's of $G$ contain various UIR's
$J^{\prime\prime}$ with various multiplicities.  This means that
we have a family of Clebsch-Gordan coefficients carrying three
sets of $J-M$ labels and in addition a multiplicity index
$\lambda$, say; and they obey two sets of unitarity conditions:
\begin{eqnarray} \sum\limits_{M,M^{\prime}} C^{J J^{\prime}
J^{\prime\prime}\lambda^{*}} _{M M^{\prime} M^{\prime\prime}}
C^{J J^{\prime} J^{\prime\prime\prime} \lambda^{\prime}}_{M
M^{\prime}M^{\prime\prime\prime}} &=&
\delta_{J^{\prime\prime}J^{\prime\prime\prime}}
\delta_{\lambda\lambda^{\prime}}
\delta_{M^{\prime\prime}M^{\prime\prime\prime}}
,\nonumber\\ \sum\limits_{J^{\prime\prime}\lambda
M^{\prime\prime}} C^{J
J^{\prime}J^{\prime\prime}\lambda^{*}} _{M M^{\prime}
M^{\prime\prime}} C^{J J^{\prime} J^{\prime\prime} \lambda}_{N
N^{\prime}M^{\prime\prime}} &=&
\delta_{MN}\delta_{M^{\prime}N^{\prime}}.  \end{eqnarray}

\noindent
Using these coefficients the product of two ${\cal D}$-functions
decomposes into a sum: 
\begin{eqnarray} {\cal D}^J_{MN}(g)
{\cal D}^{J^{\prime}}_{M^{\prime}N^{\prime}}(g) = \sum\limits
_{J^{\prime\prime}\lambda M^{\prime\prime} N^{\prime\prime}}
C^{J J^{\prime} J^{\prime\prime}\lambda^{*}}_{M M^{\prime}
M^{\prime\prime}}  C^{J J^{\prime}J^{\prime\prime} \lambda}_{N
N^{\prime}  N^{\prime\prime}} {\cal D}^{J^{\prime\prime}}
_{M^{\prime\prime} N^{\prime\prime}}(g) .  
\label{A.27}
\end{eqnarray}

\noindent
In all these relations the multiplicity index $\lambda$
accompanying the ``final'' UIR $J^{\prime\prime}$  runs over as
many values as the number of times  $J^{\prime\prime}$ occurs in
the product of $J$ and $J^{\prime}$; and at each stage we have
manifest unitary invariance under changes in the choice of
$\lambda$'s.  Combining eqns.$(\ref{A.27})$ in turn with eqns.$(\ref{A.10},
\ref{A.20})$ we immediately get the results for the products of two
$U_{JMN}$'s and the action of a $U_{JMN}$ on a state
$|J^{\prime}M^{\prime}N^{\prime}\rangle$: 
\begin{eqnarray}
U_{JMN} U_{J^{\prime}M^{\prime}N^{\prime}} &=&
\sum\limits_{J^{\prime\prime}\lambda
M^{\prime\prime}N^{\prime\prime}} C^{J J^{\prime}
J^{\prime\prime}\lambda^*}_{M M^{\prime} M^{\prime\prime}} C^{J
J^{\prime} J^{\prime\prime}\lambda} _{N
N^{\prime}N^{\prime\prime}} U_{J^{\prime\prime}M^{\prime\prime}
N^{\prime\prime}} ,\nonumber\\ U_{JMN}
|J^{\prime}M^{\prime}N^{\prime}\rangle &=&
\sum\limits_{J^{\prime\prime}\lambda
M^{\prime\prime}N^{\prime\prime}}
\sqrt{\displayfrac{N_{J^{\prime}}} {N_{J^{\prime\prime}}}} C^{J
J^{\prime}J^{\prime\prime}\lambda^*} _{M
M^{\prime}M^{\prime\prime}}
C^{JJ^{\prime}J^{\prime\prime}\lambda}_{N N^{\prime}
N^{\prime\prime}}
|J^{\prime\prime}M^{\prime\prime}N^{\prime\prime}\rangle  .
\label{A.28}
\end{eqnarray}

\noindent
The unitary invariance with respect to $\lambda$ is manifest.

Thus we have expressions $(\ref{A.6},\ref{A.20})$ for the actions of
$U\ldots, V(\cdot), \tilde{V}(\cdot)$ on $|g\rangle$, and
expressions $(\ref{A.28}, \ref{A.11})$ for their actions on $|JMN\rangle$.

Lastly we consider the question of setting up in a natural way a
complete trace orthonormal set of operators on ${\cal
H}=L^2(G)$, involving the $U$'s, $V$'s and $\tilde{V}$'s in a
``symmetrical'' manner.  In the Cartesian case the phase space
displacement operators \begin{eqnarray}
e^{i(\sigma\hat{q}-\tau\hat{p})} = e^{i\sigma\hat{q}}
e^{-i\tau\hat{p}} e^{-i\sigma\tau/2} = e^{-i\tau\hat{p}}
e^{i\sigma\hat{q}} e^{i\sigma\tau/2} \end{eqnarray}

\noindent
give us such a system, and they are basic to the Weyl
correspondence.  Already in the $\widehat{\theta}-\widehat{M}$
case we know from eqn.$(\ref{25a},\ref{25b})$ that we have to work with the
operators \begin{eqnarray} e^{in\widehat{\theta}}
e^{-i\tau\widehat{M}} e^{-in\tau/2} = e^{-i\tau\widehat{M}} e^{i
n\widehat{\theta}} e^{in\tau/2} \end{eqnarray}

\noindent
which are again complete and trace orthonormal, but we can no
longer write these as single exponentials.  In the case of
general $G$, this latter trend continues.  Generalising from the
known examples, we now define a family of operators labelled by
$g\in  G$ together with $JMN$, as follows: 
\begin{eqnarray}
\widehat{{\cal D}} (g;JMN) = V(g) U_{JMN}=\sum\limits_{M^{\prime}}
{\cal D}^J_{MM^{\prime}}(g^{-1}) U_{JM^{\prime}N} V(g) .
\label{A.31}
\end{eqnarray}

\noindent
It is easy to show trace orthogonality : using eqns. $(\ref{A.6},\ref{A.20})$,
\begin{eqnarray}
\mbox{Tr}\left(\widehat{{\cal D}}(g^{\prime};J^{\prime}
M^{\prime}N^{\prime})
^{\dag}\widehat{{\cal D}}(g;JMN)\right) &=& \int\limits_{G}
dg^{\prime\prime}\langle g^{\prime\prime}|
U^{\dag}_{J^{\prime}M^{\prime}N^{\prime}} V(g^{\prime})^{-1}
V(g) U_{JMN}|g^{\prime\prime}\rangle \nonumber\\ &=&
\int\limits_{G} dg^{\prime\prime} {\cal D}^{J^{\prime}}
_{M^{\prime}N^{\prime}}(g^{\prime\prime})^{*}
{\cal D}^{J}_{MN}(g^{\prime\prime}) \langle
g^{\prime\prime}|V(g^{\prime -1}g)|g^{\prime\prime}\rangle
\nonumber\\ &=& \delta (g^{\prime -1}g) \delta_{J^{\prime}J}
\delta_{M^{\prime}M} \delta_{N^{\prime}N} /N_J .  \end{eqnarray}

\noindent
As for completeness we begin with 
\begin{eqnarray}
\widehat{{\cal D}}(g;JMN)|g^{\prime}\rangle 
= {\cal D}^J_{MN} (g^{\prime})
|gg^{\prime}\rangle , \end{eqnarray}

\noindent
multiply both sides by $N_J {\cal D}^J_{MN}(g^{\prime\prime})^*$, 
sum on $JMN$ and use eqn.$(\ref{A.9c})$ to get: 
\begin{eqnarray}
\sum\limits_{JMN} N_J {\cal D}^J_{MN}(g^{\prime\prime})^* 
\widehat{{\cal D}}
(g; JMN)|g^{\prime}\rangle &=& \delta\left(g^{\prime\prime
-1}g^{\prime}\right)|gg^{\prime}\rangle \nonumber\\ &=&
|gg^{\prime\prime}\rangle\langle
g^{\prime\prime}|g^{\prime}\rangle .  
\end{eqnarray}

\noindent
Peeling off $|g^{\prime}\rangle$ and then replacing
$gg^{\prime\prime}$ by $g^{\prime}$ we get: 
\begin{eqnarray}
|g^{\prime}\rangle\langle g^{\prime\prime}| = \sum\limits_{JMN}
N_J {\cal D}^J_{MN} (g^{\prime\prime})^* \widehat{{\cal D}}
\left(g^{\prime}
g^{\prime\prime -1}; JMN\right) .  \end{eqnarray}

\noindent
This shows, albeit in a somewhat formal manner, that any
operator on ${\cal H}$ can be linearly expanded in the set
$\widehat{{\cal D}}(g; JMN)$.  If in eqn.$(\ref{A.31})$ we use
$\tilde{V}(\cdot)$ in place of $V(\cdot)$ we get the alternative
results: 
 \begin{eqnarray}
\widehat{\tilde{{\cal D}}}(g;JMN) = \tilde{V}(g) U_{JMN}&=&
\sum\limits_{N^{\prime}} {\cal D}^J_{N^{\prime}N} (g) 
U_{JMN^{\prime}}
\tilde{V}(g) ;\label{A.36a}\\
\mbox{Tr}\left(\widehat{\tilde{{\cal D}}}(g^{\prime};
J^{\prime}M^{\prime}
N^{\prime})^{\dag} \widehat{\tilde{{\cal D}}}(g; JMN)\right) &=&
\delta(g^{\prime -1} g)~ \delta_{J^{\prime}J}\delta_{M^{\prime}M}
\delta_{N^{\prime}N}/N_J ;\label{36b}\\ |g^{\prime}\rangle\langle
g^{\prime\prime}| &=& \sum\limits_{JMN} N_J
{\cal D}^J_{MN}(g^{\prime\prime})^* \widehat{\tilde{{\cal D}}}
(g^{\prime\prime}g^{\prime -1}; JMN). 
\label{A.36c}
\end{eqnarray}

One can ask whether similar completeness statements can be
developed for outer products of vectors of the form
$|JMN\rangle\langle J^{\prime}M^{\prime}N^{\prime}|$.  This is
indeed possible, but the expressions are somewhat unwieldy and
involve the Clebsch-Gordan coefficients explicity, so we omit
them.

The results $(\ref{A.36a}-\ref{A.36c})$ prove that the 
representation of $\widehat{{\cal A}}, V(\cdot)$ and 
$\tilde{V}(\cdot)$ on ${\cal H}=L^2(G)$ is irreducible, since any 
operator on ${\cal H}$ is expressible as a linear combination of 
the operators $\widehat{{\cal D}}(g;JMN)$ $\left(\mbox{or}\;
\widehat{\tilde{{\cal D}}}(g;JMN)\right)$.

\newpage
\def\theequation{B.\arabic{equation}}
\appendix{\bf{Appendix B}}
\setcounter{equation}{0}
\noindent

Here we briefly outline the proofs for eqns.$(\ref{78a},\ref{81})$ and also
derive some useful relations similar in form to those known in the Cartesian
and angle-angular momentum cases.

To prove $(\ref{78a})$, consider its  LHS : 
\begin{equation}
\sum_{JMM^\prime} N_{J}^{-1}\int dg {\tilde W}_1(g; JMM^\prime)
{\tilde W}_2(g; JM^\prime M). 
\label{B.1}
\end{equation} 
On substituting for ${\tilde W}$ using $(\ref{78})$ and carrying out the
summation over $JMM^\prime$ using $(\ref{A.9c})$, this expression becomes 
 
\begin{eqnarray} 
\int dg&&\int dg_{1}^{\prime} \int dg_{1}^{\prime\prime}
\int dg_{2}^{\prime} \int dg_{2}^{\prime\prime}
<g_{1}^{\prime\prime}|{\hat \rho}_1|g_{1}^{\prime}>
<g_{2}^{\prime\prime}|{\hat \rho}_2|g_{2}^{\prime}>\nonumber\\
&&~\delta(g^{-1}s(g_1^{\prime},g_{1}^{\prime\prime}))
~\delta(g^{-1}s(g_2^{\prime},g_{2}^{\prime\prime}))
~\delta(g_{1}^{\prime\prime} {g_1^{\prime}}^{-1} 
g_{2}^{\prime\prime} {g_2^{\prime}}^{-1}).
\label{B.2}
\end{eqnarray}
Using the fact that $\delta(gg^\prime)=\delta(g^\prime g)$, the third  delta 
function in the integrand can be written as $\delta( {g_1^{\prime}}^{-1} 
g_{2}^{\prime\prime} {g_2^{\prime}}^{-1}g_{1}^{\prime\prime})$ or as 
$\delta(({g_{2}^{\prime\prime}}^{-1}g_1^{\prime})^{-1}
{g_2^{\prime}}^{-1}g_{1}^{\prime\prime})$  
which in turn implies that the
integral vanishes unless
 $g_1^{\prime\prime}= g_{2}^{\prime} \cdot 
h; g_{1}^\prime =g_2^{\prime\prime}\cdot h, ~ h\in G$. This, together with the 
other two delta functions implies that $h=e$. The three delta functions 
above are therefore equivalent to 
$ \delta(g^{-1}s(g_1^{\prime},g_{1}^{\prime\prime}))
~\delta({g_{1}^{\prime}}^{-1} {g_{2}^{\prime\prime}}) 
~\delta({g_{1}^{\prime\prime}}^{-1}g_{2}^{\prime})$. 
On  carrying out the integrals in $(\ref{B.2})$ with the help of  
these delta functions one obtains the RHS of $(\ref{78a})$.

A similar line of argument can be used to establish the relation $(\ref{81})$. 
Next we show that, in analogy with the Cartesian and angle-angular momentum
cases, the Wigner distribution in $(\ref{77})$ corresponding to a density
operator ${\hat \rho}$ can be written in the following compact form
\begin{equation} 
W(g;JMN~ M^\prime N^\prime)
= 
{\rm Tr}[{\hat \rho}~~ {\widehat W}(g;JMN~ M^\prime N^\prime)],
\label{B.3}
\end{equation}
where the `Wigner operator' ${\widehat W}(g; JMN~M^\prime N^\prime)$ can be
expressed in terms of operators related to ${\widehat {\cal D}}(g;JMN)$ 
as follows:  
\begin{equation}
{\widehat W}(g; JMN~M^\prime N^\prime)= 
N_J
{\widehat {\cal D}}_1 (g;JMN)~\Delta 
~{\widehat {\cal D}}_{1}^{\dagger} (g;J M^\prime  N^\prime).
\end{equation}
Here 
\begin{eqnarray}
{\widehat {\cal D}}_1 (g;JMN)&=& U_{JMN}V(g),\label{B.5}\\
 &=& \sum_{M^\prime} {\cal D}_{MM^\prime}^{J}(g){\widehat {\cal D}} 
(g;JM^\prime N),\\
\Delta&=&\int dg \sum_{JMN} N_J {\cal D}_{MN}^{J}(e)^*{\widehat 
{\cal D}}_0 (g;JMN),\\
{\widehat {\cal D}}_0 (g;JMN)& =& \sum_{M^\prime} 
{\cal D}_{MM^\prime}^{J}(s_0(g))
{\widehat {\cal D}} (g;JM^\prime N), \\
                   &=& \sum_{M^\prime} {\cal D}_{MM^\prime}^{J}(s_0(g^{-1}))
{\widehat {\cal D}}_1 (g;JM^\prime N).\label{B.9}
\end{eqnarray}
Note that the operator ${\widehat {\cal D}}_0 (g;JMN)$ introduced here can be 
regarded as the analogue of 
$e^{ip{\hat q}-iq{\hat p}}\equiv e^{-iq{\hat p}}e^{ip{\hat q}} e^{ipq/2}$ or
of $e^{-i\tau\widehat{M}} e^{in\widehat{\theta}} e^{in\tau/2} $ in the 
angle-angular momentum case. 

To show $(\ref{B.3})$, we note that the RHS of $(\ref{B.3})$ can be written as 
\begin{equation}
{\rm Tr}[\rho~~ {\widehat W}(g;JMN~ M^\prime N^\prime)]
=\int dg_1 \int dg_2 
<g_2|\rho|g_1><g_1| {\widehat W}(g;JMN~M^\prime N^\prime)|g_2>.
\label{B.10}
\end{equation}
Now
\begin{eqnarray}
&&<g_1| {\widehat W}(g;JMN~M^\prime N^\prime)|g_2>=\nonumber\\ 
&&N_J\int dg_3 \int dg_4 <g_1|{\widehat {\cal D}}_1 (g;JMN)|g_3>
<g_3|\Delta|g_4> <g_4|{\widehat {\cal D}}_1 (g;JMN)|g_2>,
\label{B.11}  
\end{eqnarray}
and from the definitions $(\ref{B.5}-\ref{B.9})$ of the operators that occur
here it can easily be shown that 
\begin{eqnarray}
<g_1|{\widehat {\cal D}}_1 (g;JMN)|g_2>&=& {\cal D}_{MN}^J(g_1)
\delta(g_1(gg_2)^{-1}),\\
<g_1|{\widehat {\cal D}}_0 (g;JMN)|g_2>&=&{\cal D}_{MN}^J(s(g_1,g_2))
\delta(g_1(gg_2)^{-1}),\\
<g_1|\Delta|g_2>&=&
\int dg \sum_{JMN} N_J {\cal D}_{MN}^{J}(e)^*<g_1|{\widehat 
{\cal D}}_0 (g;JMN)|g_2>,\nonumber\\
&=& \int dg \sum_{JMN} N_J {\cal D}_{MN}^{J}(e)^* {\cal D}_{MN}^J(s(g_1,g_2))
\delta(g_1(gg_2)^{-1}),\nonumber\\
&=&\delta(s(g_1,g_2)).
\end{eqnarray}
Using these in $(\ref{B.11})$ one obtains
\begin{equation} 
<g_1| {\widehat W}(g;JMN, M^\prime N^\prime)|g_2>= N_J 
\delta(g^{-1}s(g_1,g_2)){\cal D}_{MN}^{J}(g_1){\cal D}_{M^\prime N^\prime
  }^{J}(g_2)^* ,
\end{equation}
 which when substituted in $(\ref{B.10})$ yields $(\ref{B.3})$. 

On setting $N=N^\prime$ ($M=M^\prime$) in ($\ref{B.3}$)  and summing over 
$N$ ($M$) we obtain the following formulae for the `simpler' Wigner
distributions in terms of `simpler' Wigner operators: 
\begin{eqnarray} 
{\tilde W}(g;JMM^\prime)
&=& {\rm Tr}[\rho~~ {\widehat {\tilde W}}(g;JMM^\prime)],\label{B.16}\\
{\tilde {\tilde W}}(g;JNN^\prime)
&=& {\rm Tr}[\rho~~ {\widehat {\tilde {\tilde W}}}(g;JNN^\prime)]
,\label{B.17}
\end{eqnarray}
where
\begin{eqnarray} 
{\widehat {\tilde W}}(g;JMM^\prime)&=&\sum_{N}
{\widehat W}(g;JMN~M^\prime N),\\
{\widehat {\tilde {\tilde W}}}(g;JNN^\prime)&=&\sum_{M}{\widehat W}(g;JMN~M N^\prime).
\end{eqnarray}

The relations $(\ref{B.16})$ and $(\ref{B.17})$ can be inverted with the help
of $(\ref{78a})$ and $(\ref{81})$  respectively to obtain 
\begin{eqnarray}
{\hat \rho}&=& 
\int dg \sum_{JMM^\prime } \frac{1}{N_J}~
{\tilde W}(g;JMM^\prime ) {\widehat {\tilde W}}(g;JM^\prime M),\label{B.20}\\
{\hat \rho}&=& 
\int dg \sum_{JNN^\prime } \frac{1}{N_J}~
{\tilde {\tilde W}}(g;JNN^\prime ) {\widehat {\tilde {\tilde W}}}(g;JN^\prime N).
\label{B.21}
\end{eqnarray}
 This can be seen as follows. Putting $\rho_1 \equiv \rho$ and $\rho_2 = 
|g_2><g_1|$ in $(\ref{78a})$ and using $(\ref{B.16})$ for the second Wigner 
distribution one
obtains 
\begin{equation}
<g_1|\rho|g_2>= 
\int dg \sum_{JMM^\prime} \frac{1}{N_J}~
{\tilde W}(g;JMM^\prime) <g_1|{\hat {\tilde W}}(g;JM^\prime M)|g_2>,
\end{equation}
which on peeling off $<g_1|$ and $|g_2>$ gives $(\ref{B.20})$. 
Eqn.$(\ref{B.21})$ can be derived in a similar fashion.


\begin{references}
\bibitem{1} E. P. Wigner, Phys. Rev. {\bf 40}, 749 (1932); For a 
comprehensive review see M. Hillery, R. F. O'Connell, M. O. Scully and
E. P. Wigner, Phys. Rep. {\bf 106}, 121 (1984) and also Y. S. Kim and 
M. E. Noz, {\it Phase Space Picture of Quantum Mechanics}, (World Scientific, 
Singapore, 1991); W. P. Schleich {\it Quantum Optics in Phase Space}, 
(Wiley-VCH, Weinheim, 2001). 
\bibitem{2} H. Weyl, Z. Phys. {\bf 46}, 1 (1927); {\it The Theory of Groups 
and Quantum Mechanics} (Dover, New York, 1950).
\bibitem{2a} Arvind, B. Dutta, N. Mukunda and R. Simon,  Pramana, J. Phys. 
{\bf 45}, 471 (1995). 
\bibitem{3} G. S. Agarwal, Phys. Rev. A {\bf 24}, 2889 (1981); 
 G. S. Agarwal, Phys. Rev. A {\bf 47}, 4608 (1993); 
J. P. Dowling, G. S. Agarwal and W. P. Schleich, Phys. Rev. A {\bf 49}, 
4101 (1994). 
\bibitem{4}J. C. V\'arilly and J. M. Gracia-Bond\'ia, Ann. Phys. NY, 
{\bf 190} 107 (1989). 
\bibitem{5} K. B. Wolf, Opt Commun. {\bf 132}, 343 (1996).
\bibitem{6} D. M. Kaplan and G. C. Summerfield, Phys. Rev. {\bf 187}, 
639 (1969). 
\bibitem{7} C. Fronsdal, Rep. Math. Phys. {\bf 15}, 111 (1979).
\bibitem{8} C. Moreno and P. Ortega-Navarro, Lett. Math. Phys. {\bf 7}, 
 181 (1983). 
\bibitem{9} R. Gilmore, {\it Lecture Notes in Physics}, {\bf 278}, ed. 
Y. S. Kim and W. W. Zachary,  (Springer, Berlin 1987), p 211; W-M. Zhang, 
D. H. Feng and R. Gilmore, Rev. Mod. Phys. {\bf 62}, 867 (1990). 
\bibitem{10} W. K. Wootters, Ann. Phys. NY {\bf 176}, 1 (1987).
\bibitem{11} U. Leonhardt, Phys. Rev. Lett. {\bf 74}, 4101 (1995); 
 U. Leonhardt, Phys. Rev. A {\bf 53}, 2998 (1996). 
\bibitem{12} C. Brif and A. Mann, J. Phys. A {\bf 31}, L9 (1998); 
Phys.Rev. A {\bf 59}, 971 (1999).
\bibitem{12a} N. M. Atakishiyev, S. M. Chumakov, and K. B. Wolf,
J. Math. Phys. {\bf 39}, 6247 (1998);
N. M. Nieto, N. M. Atakishiyev, S. M. Chumakov, and K. B. Wolf,
J. Phys. A {\bf 16}, 3875 (1998);
S. T. Ali, N. M. Atakishiyev, S. M. Chumakov, and K.B. Wolf,
Ann. Henri Poincare  {\bf 1}, 685 (2000);
M. A. Alonso, G.S. Pogosyan, and K. B. Wolf,
J. Math. Phys. {\bf 43}, 5857 (2002).
\bibitem{12b}
A.J. Bracken, D. Ellinas, and J.G. Wood,
{\it Group theory and quasiprobability integrals of Wigner functions},
quant-ph/0304010.
\bibitem{alpha} R. L. Stratonovich, Zh. Eksp. Teor. Fiz. {\bf 31}, 1012 (1956) 
(Engl. Transl. Sov. Phys.-JETP, {\bf 4}, 891 (1957) ).
\bibitem{beta} D. A. Dubin, M. A. Hennings and T. B. Smith, {\it Mathematical
    aspects of Weyl quantization and Phase}, (World Scientific, Singapore, 
2000).
\bibitem{HR} A. J. Hanson and T. Regge, Ann. Phys. (N.Y.), {\bf 87}, 498 
(1974).
\bibitem{13} N. Mukunda, Am. J. Phys. {\bf 47}, 182 (1979). 
\bibitem{14} R. L. Hudson, Rep. Math. {\bf 6}, 249 (1974). 
\bibitem{15} G. B. Folland and A. Sitaram, J. Fourier Anal. Appl. {\bf 3}
207 (1997); A. J. E. M. Janssen, J. Fourier Anal. Appl. {\bf 4}
723 (1998); P. Jaming, C. R. Acad. Sci {\bf 237}, 249 (1998).
\bibitem{16} For general information on Lie groups and the differential 
geometric aspects, see L. S. Pontrjagin, {\it Toplogical Groups}, $2^{nd}$ 
edition, (Gordon and Breach, New York, 1966); V. I. Arnold, 
{\it Mathematical Methods of Classical Mechanics}, (Springer, New York, 1978);
R. Gilmore, {\it Lie Groups, Lie Algebras and some of their Applications}, 
(Wiley, New York, 1974);  
N. Mukunda in {\it Gravitation, Gauge Theories and the Early Universe}, 
Fundamental Theories of Physics Vol. {\bf 29}, ed. B. R. Iyer, 
N. Mukunda and C. V. Vishveshwara, (Kluwer, Dodrecht, 1989).
\bibitem{gamma} L. C. Biedenharn and J. D. Louck, {\it Angular Momentum 
in Quantum Physics}, Encyclopedia of Mathematics and its Applications, 
ed. Gian-Carlo Rota, Vol. 8, (Addison-Wesley, Reading, Mass., 1981), p. 49.
\bibitem{delta} S. Chaturvedi and N. Mukunda, J. Math. Phys. {\bf 43},
5262, 5278 (2002).
\end{references}
\end{document}